# Symmetry Breaking for Answer Set Programming

Christian Drescher[*]

August 30, 2010

[*]Supported by the Austrian Science Fund (FWF) under grant number P20841 and the Vienna Science and Technology Fund (WWTF) under grant ICT08-020. Part of this work was performed when Christian Drescher was studying at the New University of Lisbon, and visiting NICTA Neville Roach Laboratory partially funded by the European Commision through ERASMUS MUNDUS Action 3, the Vienna University of Technology through KUWIS, and NICTA.

## Abstract

In the context of answer set programming, this work investigates symmetry detection and symmetry breaking to eliminate symmetric parts of the search space and, thereby, simplify the solution process. We contribute a reduction of symmetry detection to a graph automorphism problem which allows to extract symmetries of a logic program from the symmetries of the constructed coloured graph. The correctness of our reduction is rigorously proven. We also propose an encoding of symmetry-breaking constraints in terms of permutation cycles and use only generators in this process which implicitly represent symmetries and always with exponential compression. These ideas are formulated as preprocessing and implemented in a completely automated flow that first detects symmetries from a given answer set program, adds symmetry-breaking constraints, and can be applied to any existing answer set solver. We demonstrate computational impact on benchmarks versus direct application of the solver.

Furthermore, we explore symmetry breaking for answer set programming in two domains: first, constraint answer set programming as a novel approach to represent and solve constraint satisfaction problems, and second, distributed nonmonotonic multi-context systems. In particular, we formulate a translation-based approach to constraint answer set solving which allows for the application of our symmetry detection and symmetry breaking methods. To compare their performance with a-priori symmetry breaking techniques, we also contribute a decomposition of the global value precedence constraint that enforces domain consistency on the original constraint via the unit-propagation of an answer set solver. We prove correctness and evaluate both options in an empirical analysis. In the context of distributed nonmonotonic multi-context system, we develop an algorithm for distributed symmetry detection and also carry over symmetry-breaking constraints for distributed answer set programming.

# Contents

# 1   Introduction

Answer set programming (ASP; Baral, 2003) has been shown to be a useful approach for knowledge representation and nonmonotonic reasoning (NMR; Reiter, 1987) in various applications that include difficult combinatorial search, among them bioinformatics (Baral et al., 2004), crypto analysis (Aiello and Massacci, 2001), configuration (Soininen and Niemelä, 1999), database integration (Leone et al., 2005), diagnosis (Eiter et al., 1999), hardware design (Erdem and Wong, 2004), model checking (Heljanko and Niemelä, 2003), planning (Lifschitz, 2002), preference reasoning (Brewka and Eiter, 1996), semantic web (Eiter et al., 2008), and as a highlight among these applications the high-level control of the space shuttle (Nogueira et al., 2001). ASP combines an expressive but simple modelling language, able to encode all search problems within the first three levels of the polynomial hierarchy, with high-performance solving capacities (Drescher et al., 2008a).

However, many combinatorial search problems exhibit symmetries which can frustrate a search algorithm to fruitlessly explore independent symmetric subspaces. Various instance families, such as the *pigeon hole* problem, are known to require exponential time for resolution and backtracking algorithms (Urquhart, 1987). Indeed, state-of-the-art ASP solvers take a very long time to solve those instances (see Section 6). Once their symmetries are identified, it is possible to avoid redundant computational effort by pruning parts of the search space through symmetry breaking. Symmetry breaking also addresses post-processing: Where symmetries induce equivalence classes in the solution space, symmetric solutions can be discarded. Problems like the *all-interval series* taken from the CSPLib (Gent and Walsh, 1999) have plenty symmetric solution. However, all solutions to the original problem can be reconstructed from the answer sets under symmetry breaking.

This work breaks the problem of symmetry breaking down into two parts: (1) identifying symmetries and (2) breaking the identified symmetries. We adopt existing theoretical foundations from symmetry breaking for Boolean satisfiability (SAT; Biere et al., 2009) and present a reduction of symmetry detection for logic programs to the graph automorphism problem (GAP; McKay, 1981). For SAT, this has been proposed in (Crawford et al., 1996) and further refined in (Aloul et al., 2003a;b). Detected symmetries can then be utilized to add symmetry-breaking constraints (SBCs) to the original problem. These constraints ensure that a search engine never visits two points in the search space that are equivalent under the symmetry they represent. Unfortunately, generating all SBCs is intractable since there might be an exponential number of symmetries, but partial symmetry breaking can be done in polynomial time (assuming that the associated GAP is tractable). While Crawford et al. construct a partial symmetry tree, Aloul et al. restrict to a set of irredundant generators of the symmetric group. We followed the approach of Aloul et al. in (Drescher et al., 2010) and introduce an ASP encoding of symmetry-breaking constraints that is linear in the number of problem variables.

This thesis further extends our results and investigates symmetry breaking for answer set programming in two domains: first, constraint answer set programming as a novel approach to represent and solve constraint satisfaction problems (CSP), and second, distributed nonmonotonic multi-context systems (MCS; Dao-Tran et al., 2010).

**Constraint Answer Set Programming**

Constraint satisfaction problems are combinatorial search problems defined as a set of variables whose value must satisfy a number of requirements, i.e. constraints, and are subject to intense research. Problems that have been successfully modelled as a CSP stem from a variety of areas, for example, artificial intelligence, operations research, electrical engineering and telecommunications. There are several approaches to representing and solving constraint satisfaction problems: traditional constraint programming (CP; Dechter, 2003; Rossi et al., 2006), ASP, SAT, its extension to satisfiability modulo theories (SMT; Nieuwenhuis et al., 2006), and many more. Each has its particular strengths: for example, CP systems support global constraints, ASP systems permit recursive definitions and offer default negation, whilst SAT solvers often exploit very efficient implementations. In many applications it would often be helpful to exploit the strengths of multiple approaches. Consider the problem of timetabling at a university (Järvisalo et al., 2009). To model the problem, we need to express the mutual exclusion of events (for instance, we cannot place two events in the same room at the same time). A straightforward representation of such constraints with clauses and rules uses quadratic space. In contrast, global constraints such as *all-different* typically supported by CP systems can give a much more concise encoding. On the other hand, there are features that are hard to describe in traditional constraint programming, like the temporary unavailability of a particular room. However, this is easy to represent with nonmonotonic rules such as those used in ASP. Such rules also provide a flexible mechanism for defining new relations on the basis of existing ones. Answer set programming has been put forward as a powerful paradigm to solve constraint satisfaction problems by Niemelä (1999), which also shows that ASP embeds SAT but provides a more expressive framework from a knowledge representation point of view. Moreover, modern ASP solvers such as CLASP (Gebser et al., 2007b) have experienced dramatic improvements in their performance (Gebser et al., 2009c), offer an efficiency and scalability that in practice remain largely unmatched to date, and compete with the best SAT solvers (SAT competition).

An empirical comparison of the performance of ASP and constraint logic programming (CLP; Jaffar and Maher, 1994) systems on solving combinatorial problems conducted by Dovier et al. (2005) shows ASP encodings to be more compact, more declarative, and highly competitive. Particularly of relevance here is the fact that clause learning is known to be more general and potentially more powerful than traditional learning in constraint solvers (Katsirelos and Bacchus, 2005). However, Dovier et al.'s study also revealed shortcomings: non-Boolean constructs, like resources or functions over finite domains, in particular global constraints, are more naturally modelled and efficiently handled by CP systems. This led to the integration of answer set programming and constraint processing. In our work on constraint answer set programming systems (Drescher, 2010), we identify three different approaches: (1) integration of constraint solvers, (2) usage of additional propagators, such as aggregates, and (3) translation-based techniques.

Briefly, in a hybrid system, theory-specific solvers interact in order to compute solutions to the whole constraint model, similar to SMT. Hence, the key idea of an integrative approach is to incorporate constraint predicates into propositional formulas, and extending an ASP solver's decision engine for a more high-level proof procedure. Recent work on combining ASP with CP was conducted by Baselice et al. (2005); Mellarkod et al. (2008); Mellarkod and Gelfond (2008) and Gebser et al. (2009d). While Baselice et al.; Mellarkod et al.; Mellarkod and Gelfond all view ASP and CP solvers as blackboxes, Gebser et al. embed a CP solver into an ASP solver adding support for advanced backjumping and conflict-driven learning techniques. Balduccini (2009) and Järvisalo et al. (2009) cut ties to ad-hoc ASP and CP solvers, and principally support global constraints (without considering learning techniques). Dal Palù et al. (2009) put further emphasis on handling constraint variables with large domains, and presented a strategy which only consider parts of the model that actively contribute in supporting constraint answer sets. To conclude, each existing system has a subset of the following limitations: either they are tied to particular ASP and CP solvers, or the support for global constraints is limited, or communication between the ASP and CP solver is restricted.

Little attention is paid to constraint answer set programming through reformulation into ASP with usage of additional propagators, such as aggregates. Aggregations and other forms of set constructions have been shown to be useful extensions to ASP (Dell'Armi et al., 2003). In fact, a lack of aggregation capabilities may lead to an exponential growth in the number of rules required to model a CSP (Baral, 2003). Therefore, it is common to most ASP solvers to incorporate specialised algorithms, for instance, the treatment of cardinality constraints, and their generalisation to weight constraints (Niemelä et al., 1999). Work on a generic framework which provides an elegant treatment of such extensions was conducted by Elkabani et al. (2004) who employed external constraint propagators for their handling. However, it does not carry over to modern ASP solving technology based on conflict-driven nogood learning (CDNL; Gebser et al., 2007a). A first comprehensive approach to integrating specialised algorithms for weight constraint rules into CDNL is presented in (Gebser et al., 2009b).

In a translation-based approach all parts of the model are mapped into a single constraint language for which highly efficient off-the-shelf solvers are available. Previous work has mostly focussed on the translation of specific types of constraints to SAT. For example, pseudo-Boolean constraints (linear constraints over Boolean variables), including the special case of Boolean cardinality constraints, have been Booleanised such that a SAT solver can compete with the best existing native pseudo-Boolean solvers (Eén and Sörensson, 2006). Integer linear constraints have also been translated to SAT by transforming all constraints into primitive comparisons (Tamura et al., 2006). Although efficient, existent results have a number of limitations. First, the types of constraints dealt with are limited. Second, the techniques proposed are not necessarily compatible, thus making the translation of a heterogeneous constraint model difficult in both practice and theory. The latter is faced by Huang (2008) presenting translation techniques to SAT at language level by systematically Booleanising a general constraint language, rather than specialised constraint types. However, this comes with the price of weaker encodings in terms of propaga-tion power and loss of explicit domain knowledge and structure. It remains a difficult task to define universal SAT encodings that are both compact and enforce a strong type of consistency on the original model. Techniques for translating constraint variables and constraint propagation algorithms to ASP received little attention. A first study on introducing high-level statements for multi-valued propositions into the language of ASP was conducted by Gebser et al. (2009a).

We put forward translation-based constraint answer set solving in (Drescher and Walsh, 2010a;b), and show that this approach offers an efficient way to seamlessly combine the propagators of all constraints, through the unit-propagation of an ASP solver. In particular, queueing of propagators becomes irrelevant as all constraints are always propagated at once. Another major strength is that the unified conflict resolution framework can exploit constraint interdependencies, which may lead to faster propagation between constraints. Using this approach, we explore symmetry breaking in the context of constraint answer set programming.

Clearly, symmetry is an important aspect of modelling and solving CSP. Not only that symmetry occurs naturally in many problems, symmetry can also be introduced when we model a problem, e.g., if we name the elements in a variables domain, we introduce the possibility of permuting their order. We must deal with symmetry in CSP or we will waste much time visiting symmetric solutions, as well as parts of the search which are symmetric to already visited parts. In this work, we study the impact of our symmetry detection and symmetry-breaking techniques, but also studies more efficient methods for a particular, common type of value symmetry where the values of variables are interchangeable, e.g., if we have a coloured graph, we can generate another solution if we swap two colours.

**Distributed Nonmonotonic Multi-Context Systems**

With the rise of distributed systems in the world wide web, there has been increasing interest in formalisms that accommodate multiple, possibly distributed knowledge bases. Based on ground-breaking work by McCarthy (1987) and Giunchiglia (1992) several approaches have been proposed, most notably the propositional logic of context developed by McCarthy (1993) and McCarthy and Buvac (1998), and multi-context systems (MCS; Giunchiglia and Serafini, 1994), which have been associated with the local model semantics introduced by Ghidini and Giunchiglia (1998). Giunchiglia and Serafini have argued that MCSs constitute the most general among these formal frameworks.

Intuitively, an MCS consists of several heterogeneous theories (the contexts), heterogeneous in the sense that they can use different logical languages and different inference systems, that are interlinked with a special type of rules that allow to add knowledge into a context depending on knowledge in other contexts. MCSs have applications in various areas, such as argumentation, data integration, or multi-agent systems. In the latter, each context models the beliefs of an agent while the bridge rules model an agent's perception of the environment. An example for data integration from different sources is given in (Eiter et al., 2010).

Among the various MCS proposals, e.g. (Brewka et al., 2007), the general MCS framework of Brewka and Eiter (2007) is of special interest, as it generalises previous approaches in contextual reasoning and allows for heterogeneous and nonmonotonic MCSs, i.e., a system may have different, possibly nonmonotonic logics in its contexts, e.g. logic programs under answer set semantics, and bridge rules may use default negation to deal, for instance, with incomplete information. Recent work on Brewka and Eiter' style nonmonotonic MCSs was conducted by Dao-Tran et al. (2010) and Bairakdar et al. (2010). While Dao-Tran et al. assume that the topology of the MCS is not known at context nodes, Bairakdar et al. provide enhancements by computing topology information.

However, to our knowledge, symmetry breaking in MCSs is still an open issue. We instantiate the MCS framework with ASP contexts, and extend our symmetry detection and symmetry-breaking techniques to this new line of research.

**Contribution of this Thesis**

Our work addresses solving combinatorial problems in answer set programming, constraint answer set programming, and distributed answer set programming. The main contributions, briefly summarized, are as follows:

1. We present a reduction of symmetry detection in logic programs to a graph automorphism problem.

2. We propose an encoding of symmetry-breaking constraints.

3. We formulate a translation-based approach to constraint answer set solving.

4. Our decomposition of the *global value precedence* constraint enforces domain consistency.

5. We develop an algorithm for distributed symmetry detection and define distributed symmetry-breaking constraints.

6. Experimental results show the impact of symmetry breaking.

More precisely, we study and completely automate a flow that starts with a logic program and finds all of its symmetries within a very general class, including all *syntactic* symmetries, i.e., permutations that do not change the logic program. In our flow, all symmetries are captured implicitly, in terms of irredundant group generators, which always guarantees exponential compression. The logic program is then preprocessed by adding symmetry-breaking constraints that do not affect the existence of answer sets. Any ASP solver can be applied to the preprocessed logic program without changing its code, which allows for programmers to select the solvers that best fit their needs.

We contribute a reduction of symmetry detection to a graph automorphism which allows to extract symmetries of a logic program from the symmetries of the constructed graph, and also propose a construction of constraints to break detected symmetries. Experiments demonstrate computational impact. Furthermore, we extend our methods to constraint answer set programming and distributed nonmonotonic multi-context systems with ASP logics. In particular, we formulate a translation-based approach to constraint answer set solving which allows for the application of our symmetry detection and symmetry-breaking methods. To compare their performance with a-priori symmetry-breaking techniques, we also contribute a decomposition of the *global value precedence* constraint that enforces domain consistency on the original constraint using the unit-propagation of an ASP solver. We evaluate both options in an empirical analysis. In the context of distributed nonmonotonic multi-context system, we develop an algorithm for distributed symmetry detection and also carry over symmetry-breaking constraints for distributed answer set programming.

**Organisation of this Thesis**

The remaining material is organised as follows. At first, we provide all necessary preliminaries in Section 2. In particular, we describe answer set programming, our approach to constraint answer set programming and distributed nonmonotonic multi-context systems. We then give group theoretic background and define what we mean by a symmetry in Section 3. In Section 4, we present our symmetry detection techniques for logic programs, for their extensions, and for distributed nonmonotonic multi-context systems with ASP logics. Section 5 amounts to symmetry-breaking methods. We implemented our techniques in various systems, which we evaluate in Section 6. Section 7 concludes our work.

Throughout this thesis, we state theorems and provide their proofs if not proved elsewhere, and sometimes with help of lemmas. Corollaries follow from theorems and are as tagged as such. To enhance readability, we do not tag every definition, but instead, *emphasise* whenever a new term is introduced.

# 2 Logical Background

We start with an introduction to answer set programming and provide all necessary background which allows us to characterise inference in ASP as unit-propagation on nogoods. In turn, we explain our translation-based techniques for constraint answer set programming, an important extension of ASP. Running examples, most notably the *all-interval series* and the *graph colouring* problem, complement the theory. Finally, towards distributed answer set programming, we present distributed nonmonotonic multi-context systems.

## 2.1 Answer Set Programming

As a form of declarative programming oriented towards combinatorial search problems, ASP comes with an expressive but simple modelling language.

**Definition 2.1.** *A* (disjunctive) logic program *over a set of primitive propositions $\mathcal{A}$ is a finite set of* rules $r$ *of the form*

$$a_1; \ldots; a_\ell \leftarrow b_1, \ldots, b_m, \sim c_1, \ldots, \sim c_n \qquad (2.1)$$

*where $a_i, b_j, c_k \in \mathcal{A}$ are* atoms *for $1 \leq i \leq \ell$, $1 \leq j \leq m$, and $1 \leq k \leq n$.*

A *default literal* $\hat{a}$ is an atom $a$ or its default negation $\sim a$. Let $head(r) = \{a_1, \ldots, a_\ell\}$ be the *head* of $r$ and $body(r) = \{b_1, \ldots, b_m, \sim c_1, \ldots, \sim c_n\}$ the *body* of $r$. For a set $S$ of atoms, define $\overline{S} = \{\sim a \mid a \in S\}$. For a set $S$ of default literals, define $S^+ = \{a \mid a \in S\}$ and $S^- = \{a \mid \sim a \in S\}$. The set of atoms occurring in a logic program $P$ is denoted by $atom(P)$, and the set of bodies in $P$ is $body(P) = \{body(r) \mid r \in P\}$. If $|head(r)| = 1$ for all $r \in P$, i.e., all rules in the $P$ have a single head atom, we call $P$ a *normal* logic program.

The semantics of a logic program is given by its answer sets. A set $M \subseteq \mathcal{A}$ is an *answer set* of a logic program $P$ over $\mathcal{A}$, if $M$ is a $\subseteq$-minimal model of the *reduct* (Gelfond and Lifschitz, 1991)

$$P^M = \{head(r) \leftarrow body(r)^+ \mid r \in P,\ body(r)^- \cap M = \emptyset\}.$$

A rule of form (2.1) can be seen as a constraint on the answer sets of a program, stating that if $b_1, \ldots, b_m$ are in the answer set and none of $c_1, \ldots, c_n$ are included, then one of $a_1, \ldots, a_\ell$ must be in the set. Important extensions to logic programs are integrity constraints, choice rules, and cardinality constraints (Simons et al., 2002).

**Definition 2.2.** *Given an alphabet $\mathcal{A}$. An* integrity constraint *has the form*

$$\leftarrow b_1, \ldots, b_m, \sim c_1, \ldots, \sim c_n \tag{2.2}$$

*where $b_j, c_k \in \mathcal{A}$, for $1 \leq j \leq m$ and $1 \leq k \leq n$.*

We understand an integrity constraint as a short hand for a rule with an unsatisfiable head, and thus forbids its body to be satisfied in any answer set.

**Example 2.1.** *Consider the logic programs $P_1$ and $P_2$, both have two answer sets $\{a\}$ and $\{b\}$, given by*

$$P_1 = \left\{ \begin{array}{c} a \leftarrow \sim b \\ b \leftarrow \sim a \end{array} \right\}, \qquad P_2 = \left\{ \begin{array}{c} a;b \leftarrow \\ \leftarrow a,b \end{array} \right\}.$$

*To verify, for instance, answer set $\{a\}$, we consider the reduct $P_1^{\{a\}}$, $P_2^{\{a\}}$ respectively:*

$$P_1^{\{a\}} = \{\ a \leftarrow\ \}, \qquad P_2^{\{a\}} = \{\ a;b \leftarrow\ \}.$$

*The $\subseteq$-minimal model of $P_1^{\{a\}}$ is $\{a\}$. $P_2^{\{a\}}$ has three classical models, $\{a\}$, $\{b\}$, and $\{a,b\}$ where $\{a\}$ and $\{b\}$ are $\subseteq$-minimal. Therefore, $\{a\}$ is an answer set of both $P_1$ and $P_2$. Observe that $P_1$ and $P_2$ remain invariant under a swap of atoms $a$ and $b$, which is what we call a symmetry. In this work we will only deal with symmetries that can be thought of as permutations of atoms.*

**Definition 2.3.** *Given an alphabet $\mathcal{A}$. An* choice rule *has the form*

$$\{a_1, \ldots, a_\ell\} \leftarrow b_1, \ldots, b_m, \sim c_1, \ldots, \sim c_n \ . \tag{2.3}$$

*where $a_i, b_j, c_k \in \mathcal{A}$, for $1 \leq i \leq \ell$, $1 \leq j \leq m$, and $1 \leq k \leq n$.*

A choice rule allows for the nondeterministic choice over atoms in $\{a_1, \ldots, a_n\}$.

**Definition 2.4.** *Given an alphabet $\mathcal{A}$. A* cardinality constraint *has the form*

$$\leftarrow k\{\hat{a}_1, \ldots, \hat{a}_n\} \tag{2.4}$$

*where $a_i \in \mathcal{A}$, for $1 \leq i \leq n$ and $k \geq 0$ is an integer.*

A cardinality constraint is interpreted as no answer set satisfies $k$ or more default literals of the set $\{\hat{a}_1, \ldots, \hat{a}_n\}$.

More formally, the semantics of integrity constraints, choice rules, and cardinality constraints can be given through program transformations that introduce additional propositions (Simons et al., 2002). For instance, a cardinality constraint of the form (2.4) can be transformed into $\binom{n}{k}$ integrity constraints $r$ such that $body(r) \subseteq \{\hat{a}_1, \ldots, \hat{a}_n\}$ and $|body(r)| = k$. Simons et al. provide a transformation that needs just $\mathcal{O}(nk)$ rules, introducing atoms $l(\hat{a}_i, j)$ to represent the fact that at least $j$ of the default literals with index $\geq i$, i.e., the default literals in $\{\hat{a}_i, \ldots, \hat{a}_n\}$, are in a particular answer set candidate. Then, the cardinality constraint can be encoded by an integrity constraint $\leftarrow l(\hat{a}_1, k)$ and the three following rules, where $1 \leq i \leq n$ and $1 \leq j \leq k$:

$$\begin{aligned} l(\hat{a}_i, j) &\leftarrow l(\hat{a}_{i+1}, j) \\ l(\hat{a}_i, j+1) &\leftarrow \hat{a}_i, l(\hat{a}_{i+1}, j) \\ l(\hat{a}_i, 1) &\leftarrow \hat{a}_i \end{aligned}$$

Notice that both transformations are modular. Alternatively, modern ASP solvers also incorporate specialised propagators for cardinality constraints that run in $\mathcal{O}(n)$.

Although the answer set semantics are propositional, atoms in $\mathcal{A}$ and can be constructed from a first-order signature $\Sigma_\mathcal{A} = (\mathcal{F}_\mathcal{A}, \mathcal{V}_\mathcal{A}, \mathcal{P}_\mathcal{A})$, where

- $\mathcal{F}_\mathcal{A}$ is a set of function symbols (including constant symbols),

- $\mathcal{V}_\mathcal{A}$ is a denumerable collection of (first-order) variables, and

- $\mathcal{P}_\mathcal{A}$ is a set of predicate symbols.

The logic program over $\mathcal{A}$ is then obtained by a *grounding* process, systematically substituting all occurrences of variables $\mathcal{V}_\mathcal{A}$ by terms in $\mathcal{T}(\mathcal{F}_\mathcal{A})$, where $\mathcal{T}(\mathcal{F}_\mathcal{A})$ denotes the set of all ground terms over $\mathcal{F}_\mathcal{A}$. Atoms in $\mathcal{A}$ are formed from predicate symbols $\mathcal{P}_\mathcal{A}$ and terms in $\mathcal{T}(\mathcal{F}_\mathcal{A})$.

ASP engineers usually use a *generate-and-test* technique (Baral, 2003) to model a problem, by producing the space of solution candidates in the *generate* component and defining rules that filter invalid solutions in the *test* component. For instance, the first line of our *all-interval series* problem encoding from Examples 2.2 generates an assignment to the problem variables. The remaining rules comprise the *test* component as they eliminate assignments that do not solve the problem.

**Example 2.2.** *The* all-interval series *problem is to find a permutation of the $n$ integers from $0$ to $n-1$ such that the difference of adjacent numbers are also all-different. We encode the* all-interval series *problem introducing propositional variables $v_{i,j}$ and $d_{k,l}$ for the $i \in 1..n$ integer variables taking values $0 \leq j < n$, and $k \in 1..(n-1)$ auxiliary variables taking values $l \in 1..(n-1)$ to represent the differences between adjacent numbers, respectively. Furthermore, we require both sets of variables to have pairwise different values.*

$$\begin{aligned} v_{i,0}; v_{i,1}; \ldots; v_{i,n-1} &\leftarrow & i &\in 1\ldots n \\ &\leftarrow v_{i,k}, v_{j,k} & i &< j \\ d_{i,|j-k|} &\leftarrow v_{i,j}, v_{i+1,k} & i &\in 1..(n-1) \wedge j,k \in 0\ldots(n-1) \\ &\leftarrow d_{i,k}, d_{j,k} & i &< j \end{aligned}$$

*Note that above encoding remains invariant under complex permutation of atoms. We refer to Example 3.3 for a detailed analysis.*

Generate-and-test is also used in the encoding of *Ramsey's theorem* in Section 6.2.

**Nogoods**

As shown by Lee (2005), the answer sets of a logic program $P$ correspond to the classical models of $P$ that satisfy all loop formulas, where the classical models of $P$ are represented by the set of formulas

$$RF_P = \left\{ \left( \bigwedge_{b \in body(r)^+} b \wedge \bigwedge_{c \in body(r)^-} \neg c \right) \rightarrow \bigvee_{a \in head(r)} a \mid r \in P \right\},$$

A nonempty set $L \subseteq \mathcal{A}$ is called a *loop* of $P$, if for all nonempty $K \subset L$, there is some $r \in P$ such that $head(r) \cap K \neq \emptyset$ and $body(r) \cap (L \setminus K) \neq \emptyset$ (Gebser et al., 2006). Note that every atom contained in $\mathcal{A}$, forms a loop of $P$, i.e. a singleton, and if all loops are singletons, then $P$ is called *tight* (Erdem and Lifschitz, 2003). For a loop $L$, let

$$supp_P(L) = \{r \in P \mid head(r) \cap L \neq \emptyset,\ body(r) \cap L = \emptyset\}\ .$$

be the set of rules from $P$ that can externally support $L$. The *(disjunctive) loop formula* (Lee, 2005) of $L$ is defined as

$$LF_P(L) = \bigvee_{a \in L} a \rightarrow \bigvee_{r \in supp_P(L)} \left( \bigwedge_{b \in body(r)^+} b \wedge \bigwedge_{c \in body(r)^-} \neg c \wedge \bigwedge_{d \in head(r) \setminus L} \neg d \right).$$

Finally, let $loop(P)$ denote the set of all loops in $P$ and $LF_P = \{LF_P(L) \mid loop(P)\}$. Then, according to Lee, a set $M \subseteq \mathcal{A}$ is an answer set of a logic program $P$, if $M$ is a model of $RF_P \cup LF_P$.

As a remark, the influential Clark's completion (Clark, 1978) allows for a slightly different characterisation, where the *completion* $Comp(P)$ *of a logic program* $P$ is defined as the set of rules in $P$, and the loop formulas for singletons. Hence, a set $M \subseteq \mathcal{A}$ is an answer set of a logic program $P$, if $M$ is a model of $Comp(P) \cup LF_P$ (Ben-Eliyahu and Dechter, 1994; Lee and Lifschitz, 2003).

We want to view inferences in ASP as unit-propagation on nogoods. Following Gebser et al. (2007a), inferences in ASP rely on atoms and program rules, which can be expressed by using atoms and bodies. Thus, for a program $P$, the *domain* of truth assignments $\mathbf{A}$ is fixed to $dom(\mathbf{A}) = atom(P) \cup body(P)$. Formally, a *truth assignment* $\mathbf{A}$ is a set $\{\sigma_1, \ldots, \sigma_n\}$ of *signed literals* $\sigma_i$ for $1 \leq i \leq n$ of the form $\mathbf{T}a$ or $\mathbf{F}a$ where $a \in dom(\mathbf{A})$. $\mathbf{T}a$ expresses that $a$ is assigned *true* and $\mathbf{F}a$ that it is *false* in $\mathbf{A}$. (We omit the attribute *truth* for assignments whenever clear from the context.) The complement of a signed literal $\sigma$ is denoted by $\overline{\sigma}$, that is $\overline{\mathbf{T}a} = \mathbf{F}a$ and $\overline{\mathbf{F}a} = \mathbf{T}a$. In the context of ASP, we define a *nogood* (Dechter, 2003) as follows.

**Definition 2.5.** *A* nogood *is a set* $\delta = \{\sigma_1, \ldots, \sigma_n\}$ *of signed literals, expressing a constraint violated by any assignment* $\mathbf{A}$ *such that* $\delta \subseteq \mathbf{A}$.

For a nogood $\delta$, a signed literal $\sigma \in \delta$, and an assignment $\mathbf{A}$, we say that $\delta$ is *unit* and $\overline{\sigma}$ is *unit-resulting* if $\delta \setminus \mathbf{A} = \{\sigma\}$. Let $\mathbf{A^T} = \{a \in dom(\mathbf{A}) \mid \mathbf{T}a \in A\}$ the set of true propositions and $\mathbf{A^F} = \{a \in dom(\mathbf{A}) \mid \mathbf{F}a \in A\}$ the set of false propositions. A *total* assignment, that is $\mathbf{A}^\mathbf{T} \cup \mathbf{A}^\mathbf{F} = dom(\mathbf{A})$ and $\mathbf{A}^\mathbf{T} \cap \mathbf{A}^\mathbf{F} = \emptyset$, is a *solution* for a set $\Delta$ of nogoods if $\delta \not\subseteq \mathbf{A}$ for all $\delta \in \Delta$. This provides us with a uniform framework for describing propagation via a program, its completion, and loop formulas.

We follow Drescher et al. (2008a) and start by considering a logic program $P$ as sets of implications, i.e., we represent $RF_P$. For a set $S = \{b_1, \ldots, b_m, \sim c_1, \ldots, \sim c_n\}$ of default literals, the following set $\Delta(S)$ of nogoods defines whether $S$ must be assigned $\mathbf{T}$ or $\mathbf{F}$ in terms of the conjunction its elements:

$$\Delta(S) = \left\{ \begin{array}{l} \{\mathbf{T}b_1, \ldots, \mathbf{T}b_m, \mathbf{F}c_1, \ldots \mathbf{F}c_n, \mathbf{F}S\}, \\ \{\mathbf{F}b_1, \mathbf{T}S\}, \ldots, \{\mathbf{F}b_m, \mathbf{T}S\}, \{\mathbf{T}c_1, \mathbf{T}S\}, \ldots, \{\mathbf{T}c_n, \mathbf{T}S\} \end{array} \right\}.$$

Intuitively, for some rule $r$, the nogoods in $\Delta(body(r))$ enforce the truth of $body(r)$ iff all its default literals are satisfied. This allows us to characterise the implications expressed by a program $P$ via the following set of nogoods:

$$\Delta_P = \bigcup_{r \in P} (\Delta(body(r)) \cup \{\{\mathbf{T}body(r)\} \cup \{\mathbf{F}a \mid a \in head(r)\}\}) .$$

Then, the solutions for $\Delta_P$ correspond to the classical models of $P$. In order to describe the completion of a logic program $P$ via nogoods, we make use of *shifting* (Gelfond et al., 1991):

$$\vec{P} = \{a_i \leftarrow body(r), \sim a_1, \ldots, \sim a_{i-1}, \sim a_{i+1}, \ldots \sim a_\ell \mid r \in P,$$
$$head(r) = \{a_1, \ldots, a_{i-1}, \sim a_i, \sim a_{i+1}, \ldots, \sim a_\ell\}\} .$$

Shifting retains the loop formulas for singletons, what allows us to check support of singletons on the shifted version of a program (Drescher et al., 2008a). The set of nogoods

$$\Theta_{\vec{P}} = \bigcup_{r \in \vec{P}} \Delta(body(r)) \cup \{\{\mathbf{T}a\} \cup \{\mathbf{F}d \mid d \in body(sup_{\vec{P}}(\{a\}))\} \mid a \in atom(P)\}$$

then regulate support for singletons. Given that for a tight program $P$ every loop of $P$ is a singleton, the solutions for $\Delta_P \cup \Theta_{\vec{P}}$ coincide with the models of $RF_P \cup LF_P$, i.e., the answer sets of $P$. This result still holds after replacing $\Delta_P$ by $\Delta_{\vec{P}}$. Note that $\Delta_P \cup \Theta_{\vec{P}}$ amounts to the completion of $P$, provided that $P$ does not contain any tautological rules $r$, i.e., $head(r) \cap body(r) \neq \emptyset$. Loop formulas, expressed in the set of nogoods $\Lambda_P$, have to be added to establish full correspondence to the answer sets of $P$. We refer the reader to (Drescher et al., 2008a) for details. Typically, solutions for $\Delta_P \cup \Theta_{\vec{P}} \cup \Lambda_P$ are computed by applying some form of conflict-driven nogood learning (Gebser et al., 2007a; Drescher et al., 2008b). This combines search and propagation by recursively assigning the value of a proposition and using *unit-propagation* to determine logical consequences (Mitchell, 2005).

**Example 2.3.** *Reconsider the logic program $P_1$ from Example 2.1. We have*

$$\Delta_{P_1} = \{\{\mathbf{F}b, \mathbf{F}\beta_1\}, \{\mathbf{T}b, \mathbf{T}\beta_1\}, \{\mathbf{F}a, \mathbf{F}\beta_2\}, \{\mathbf{T}a, \mathbf{T}\beta_2\}, \{\mathbf{T}\beta_1, \mathbf{F}a\}, \{\mathbf{F}\beta_2, \mathbf{T}b\}\}$$
$$\Theta_{\vec{P}_1} = \{\{\mathbf{T}a, \mathbf{F}\beta_1\}, \{\mathbf{F}b, \mathbf{T}\beta_2\}\}$$

*Since $P_1$ is tight, $\Delta_P \cup \Theta_{\vec{P}_1}$ coincide with the answer sets of $P_1$. Suppose $\mathbf{A} = \{\mathbf{F}b\}$. Then the nogoods $\{\mathbf{F}b, \mathbf{F}\beta_1\}$ and $\{\mathbf{F}b, \mathbf{T}\beta_2\}$ are unit, where $\mathbf{F}\beta_1$ and $\mathbf{T}\beta_2$ are unit-resulting and trigger further propagation. Eventually, unit-propagation extends $\mathbf{A}$ to $\{\mathbf{F}b, \mathbf{F}\beta_1, \mathbf{T}\beta_2, \mathbf{T}a\}$, that is the answer set $\{a\}$.*

```
Algorithm:  UP(∇, A)
Input:      A set ∇ of nogoods, and an assignment A.
Output:     An extended assignment, and a status (either violating or success).

            repeat
                if δ ⊆ A for some δ ∈ ∇ then
                    return (A, violating);
                Σ ← {δ ∈ ∇ | δ \ A = {σ}, σ̄ ∉ A};
                if Σ ≠ ∅ then let σ ∈ δ \ A for some δ ∈ Σ in
                    A ← A ∪ (σ̄);
            until Σ = ∅;
            return (A, success);
```

Algorithm 2.1: The unit-propagation algorithm.

## 2.2 Constraint Answer Set Programming

The classic definition of a constraint satisfaction problem is as follows (Rossi et al., 2006).

**Definition 2.6.** *A* constraint satisfaction problem *is a triple* $(V, D, C)$ *where $V$ is a set of* variables $V = \{v_1, \ldots, v_n\}$, *$D$ is a set of finite* domains $D = \{D_1, \ldots, D_n\}$ *such that each variable $v_i$ has an associated domain $dom(v_i) = D_i$, and $C$ is a set of* constraints. *A constraint $c$ is a pair $(R_S, S)$ where $R_S$ is a $k$-ary* relation *on the variables in $S \in V^k$, called the* scope *of $c$.*

In other words, $R_S$ is a subset of the Cartesian product of the domains of the variables in $S$. To access the relation and the scope of $c$ define $range(c) = R_S$ and $scope(c) = S$. For a *(constraint variable) assignment* $A : V \to \bigcup_{v \in V} dom(v)$ such that $A(v) \in dom(v)$ for all $v \in V$, and a constraint $c = (R_S, S)$ with $S = (v_1, \ldots, v_k)$, define $A(S) = (A(v_1), \ldots, A(v_k))$, and call $c$ *satisfied* if $A(S) \in range(c)$. Given this, define the set of constraints satisfied by $A$ as $sat_C(A) = \{c \in C \mid A(scope(c)) \in range(c)\}$. A binary constraint $c$ has $|scope(c)| = 2$, while a *global* (or $n$-ary) constraint $c$ has parametrized scope.

**Example 2.4.** *The binary constraint $v_1 \neq v_2$ ensures that $v_1$ and $v_2$ take different values, while the* global all-different *constraint ensures that a set of variables, $\{v_1, \ldots, v_n\}$ take all different values. This can be decomposed into $O(n^2)$ binary constraints $v_i \neq v_j$ for $i < j$.*

Finally, an assignment $A$ is a *solution* for a CSP iff it satisfies all constraints in $C$.

**Constraint Programming**

Constraint programming is a natural paradigm for solving constraint satisfaction problems. CP systems usually use a *constrain-and-generate* technique in which an initial deterministic phase assigns a domain to each of the constraint variables and imposes a number of constraints, then a nondeterministic phase generates and explores the solution space. Various heuristics affecting, for instance, the variable selection criteria and the ordering of the attempted values, can be used to guide the search. Each time a variable is assigned a value, a deterministic propagation stage is executed, pruning the set of values to be attempted for the other variables, i.e., enforcing a certain type of *local consistency* like arc, domain, bound, or range consistency.

**Definition 2.7.** *A binary constraint c is called* arc consistent *iff when a variable $v_1 \in scope(c)$ is assigned any value $d_1 \in dom(v_1)$, there exists a consistent value $d_2 \in dom(v_2)$ for the other variable $v_2$.*

**Definition 2.8.** *An n-ary constraint c is* generalised arc consistent *(GAC) or* domain consistent *iff when a variable $v_i \in scope(c)$ is assigned any value $d_i \in dom(v_i)$, there exist compatible values in the domains of all the other variables $d_j \in dom(v_j)$ for all $1 \leq j \leq n$, $j \neq i$ such that $(d_1, \ldots, d_n) \in range(c)$.*

The concepts of bound and range consistency are defined for constraints on ordered intervals. Let $min(D_i)$ and $max(D_i)$ be the minimum value and maximum value of the domain $D_i$. A constraint $c$ is *bound consistent* iff when a variable $v_i$ is assigned either $min(dom(v_i))$ or $max(dom(v_i))$, i.e., the minimum or maximum value in its domain, there exist compatible values between the minimum and maximum domain value for all the other variables in the scope of the constraint. Such an assignment is called a *bound support*. A constraint is *range consistent* iff when a variable is assigned any value in its domain, there exists a bound support. Notice that range consistency is in between domain and bound consistency.

**Constraint Logic Programming**

Constraint logic programming is a programming paradigm that naturally merges traditional constraint programming and logic programming. The goal is to bring advantages of logic programming based knowledge representation techniques to constraint programming.

**Definition 2.9.** *A constraint logic program $P$ over an extended alphabet distinguishing regular and constraint atoms, denoted by $\mathcal{A}$ and $\mathcal{C}$, respectively, is a set of rules of the form*

$$a_1; \ldots; a_\ell \leftarrow b_1, \ldots, b_m, \sim c_1, \ldots, \sim c_n \tag{2.5}$$

*where $a_i \in \mathcal{A}$ and $b_j, c_k \in \mathcal{A} \cup \mathcal{C}$, for $1 \leq i \leq \ell$, $1 \leq j \leq m$, and $1 \leq k \leq n$.*

Observe that a constraint logic program is in fact a logic program over $\mathcal{A} \cup \mathcal{C}$. Constraint atoms are identified with constraints via a function $\gamma : \mathcal{C} \to C$. For sets of constraints, we write $\gamma(C') = \{\gamma(c) \mid c \in C'\}$ for $C' \subseteq \mathcal{C}$. Finally, for a set $S$ of default literals we define $S|_\mathcal{A} = S \cap (\mathcal{A} \cup \overline{\mathcal{A}})$ and $S|_\mathcal{C} = S \cap (\mathcal{C} \cup \overline{\mathcal{C}})$.

We extend the answer set semantics to constraint logic programs and define the *constraint reduct* (Gebser et al., 2009d) as

$$P^A = \{head(r) \leftarrow body(r)|_\mathcal{A} \mid r \in P,$$
$$\gamma(body(r)^+|_\mathcal{C}) \subseteq sat_C(A),\ \gamma(body(r)^-|_\mathcal{C}) \cap sat_C(A) = \emptyset\}.$$

Then, a set $M \subseteq \mathcal{A}$ is a *constraint answer set* of a constraint logic program $P$ with respect to an assignment $A$, if $M$ is an answer set of $P^A$.

Similar to logic programs, the atoms in $\mathcal{A}$ and $\mathcal{C}$ can be constructed from a multi-sorted, first-order signature $\Sigma = (\mathcal{F}_\mathcal{A} \cup \mathcal{F}_\mathcal{C}, \mathcal{V}_\mathcal{A} \cup \mathcal{V}_\mathcal{C}, \mathcal{P}_\mathcal{A} \cup \mathcal{P}_\mathcal{C})$, where

- $\mathcal{F}_\mathcal{A} \cup \mathcal{F}_\mathcal{C}$ is a finite set of function symbols (including constant symbols),
- $\mathcal{V}_\mathcal{A}$ is a denumerable collection of regular variable symbols,
- $\mathcal{V}_\mathcal{C} \subseteq \mathcal{T}(\mathcal{F}_\mathcal{A})$ is a set of constraint variable symbols, and
- $\mathcal{P}_\mathcal{A} \cup \mathcal{P}_\mathcal{C}$ is a finite set of predicate symbols, where $\mathcal{P}_\mathcal{A}$ and $\mathcal{P}_\mathcal{C}$ are disjoint.

While the atoms in $\mathcal{A}$ are formed as discussed before, the ones in $\mathcal{C}$ are constructed from predicate symbols $\mathcal{P}_\mathcal{C}$ and $(\mathcal{F}_\mathcal{C}, \mathcal{V}_\mathcal{C})$-terms. This definition follows Gebser et al. (2009d) and tolerates occurrences of similar ground terms in atoms of both $\mathcal{A}$ and $\mathcal{C}$.

**Example 2.5.** *To illustrate constraint answer set programming, we encode the* graph colouring *problem in the language of the preprocessor* INCA *from the (Potassco suite) labs suite. A colouring of a graph $(V, E)$ is a mapping $c : V \to \{1, \ldots, k\}$ such that $c(v) \neq c(w)$ for every edge $(v, w) \in E$ with a given number $k$ of colours. Given $k$, the* graph colouring *problem is to determine the existence of a colouring.*

```
#var $colour(X) : node(X) = 1..k.

:- $colour(X) == $colour(Y), edge(X,Y).
```

*The first line defines an integer variable for each node, taking values from 1 to $k$, representing the colouring. The second line posts the constraint that connected nodes must not have the same colour. We have $\mathcal{F}_\mathcal{A}$ is the union of $\{\$colour\}$ and the set of all possible arguments of the edge/2-relation defined in some problem instance, $\mathcal{F}_\mathcal{C} = \emptyset$, $\mathcal{V}_\mathcal{A} = \{X, Y\}$, $\mathcal{P}_\mathcal{A} = \{edge\}$, and $\mathcal{P}_\mathcal{C} = \{==\}$. Observe that the colours are symmetric to each other (value symmetry).*

In (Drescher and Walsh, 2010a;b) we explain how to translate constraint logic programs with multi-valued propositions into a logic program. There are a number of choices of how to encode constraints on multi-valued propositions, e.g. a constraint variable $v$, taking values out of a predefined finite domain, $dom(v)$. In what follows, we assume $dom(v) = [1, d]$ for all $v \in V$ to save the reader from multiple superscripts.

A popular choice is called the *direct encoding* (Walsh, 2000). In the direct encoding, a propositional variable $e(v, i)$, representing $v = i$, is introduced for each value $i$ that can be assigned to the constraint variable $v$. Intuitively, the proposition $e(v, i)$ is true if $v$ takes the value $i$, and false if $v$ takes a value different from $i$. For each $v$, the truth-assignments of atoms $e(v, i)$ are encoded by a choice rule (2.6). Furthermore, there is an integrity constraint (2.7) to ensure that $v$ takes at least one value, and a cardinality constraint (2.8) that ensures that $v$ takes at most one value.

$$\{e(v, 1), \ldots, e(v, d)\} \leftarrow \tag{2.6}$$
$$\leftarrow \sim e(v, 1), \ldots, \sim e(v, d) \tag{2.7}$$
$$\leftarrow 2 \{e(v, 1), \ldots, e(v, d)\} \tag{2.8}$$

In the direct encoding, each forbidden combination of values in a constraint is expressed by an integrity constraint. On the other hand, when a relation is represented by allowed combinations of values, all forbidden combinations have to be deduced and translated to integrity constraints. Unfortunately, the direct encoding of constraints hinders propagation:

**Theorem 2.1** (Walsh, 2000). *Enforcing arc consistency on the binary decomposition of the original constraint prunes more values from the variables domain than unit-propagation on its direct encoding.*

The *support encoding* has been proposed to tackle this weakness (Gent, 2002). A *support* for $v = 9$ in a constraint $c$ is the set of values $\{i_1, \ldots, i_m\} \subseteq dom(v')$ of another variable in $v' \in scope(c) \setminus \{v\}$ which allow $v = i$, and can be encoded as following *support rule*, extending (2.6–2.8):

$$\leftarrow e(v, i), \sim e(v', i_1), \ldots, \sim e(v', i_m) \ .$$

This integrity constraint can be read as whenever $v = i$, then at least one of its supports must hold. In the support encoding, for each constraint $c$ there is one support rule for each pair of distinct variables $v, v' \in scope(c)$, and for each value $i$.

**Theorem 2.2** (Gent, 2002). *Unit-propagation on the support encoding enforces arc consistency on the binary decomposition of the original constraint.*

We illustrate this approach on an encoding of the *global all-different* constraint. For variables $v, v'$ and value $i$ it is defined by the following $\mathcal{O}(n^2 d)$ integrity constraints:

$$\leftarrow e(v, i), \sim e(v', 1), \ldots, \sim e(v', i-1), \sim e(v', i+1), \ldots, \sim e(v', d)$$

To keep the encoding small, we make use of the equivalence

$$e(v', i) \equiv \sim e(v', 1), \ldots, \sim e(v', i-1), \sim e(v', i+1), \ldots, \sim e(v', d) \tag{2.9}$$

covered by (2.7–2.8) and get

$$\leftarrow e(v, i), e(v', i) \ .$$

Observe that this is also the direct encoding of the binary decomposition of the *global all-different* constraint. However, this observation does not hold in general for all constraints (Walsh, 2000). As discussed in the previous section of this thesis, we can use Simons et al.'s encoding to optimise above condition, or rather express it as $\mathcal{O}(d)$ cardinality constraints:

$$\leftarrow 2 \{e(v_1, i), \ldots, e(v_n, i)\} \ . \tag{2.10}$$

**Corollary 2.3** (Drescher and Walsh, 2010a). *Unit-propagation on (2.6–2.8) and (2.10) enforces arc consistency on the binary decomposition of the global* all-different *constraint in $\mathcal{O}(nd^2)$ down any branch of the search tree.*

In (Drescher and Walsh, 2010a;b) we also propose a *range encoding* and a *bound encoding*, and prove similar results, i.e., unit-propagation on the range encoding enforces range consistency on the original constraint, and unit-propagation on the bound encoding enforces bound consistency on the original constraint. In particular, we show how simple encodings can simulate very complex propagation algorithm with a similar overall complexity of reasoning.

## 2.3 Distributed Nonmonotonic Multi-Context Systems

The idea of heterogeneous multi-context systems is to allow different logics to be used in different contexts, and to provide a framework that allows to add knowledge into a context depending on knowledge in other contexts. Following Brewka and Eiter (2007), a *logic* $L = (\mathbf{KB}, \mathbf{BS}, \mathbf{ACC})$ is composed of the following components:

- $\mathbf{KB}$ is the set of well-formed knowledge bases (sets of formulas) of $L$.
- $\mathbf{BS}$ is the set of possible belief sets (sets of formulas),
- $\mathbf{ACC} : \mathbf{KB} \to 2^{\mathbf{BS}}$ is a function describing the semantics of the logic by assigning each $kb \in \mathbf{KB}$ a set of acceptable sets of beliefs.

This covers many monotonic and nonmonotonic logics like *propositional logic* (c.f. Biere et al., 2009) under the closed world assumption and *default logic* (Reiter, 1980). We will consider normal logic programs under answer set semantics, i.e., ASP logics $L$ such that

- $\mathbf{KB}$ is the set of normal logic programs over an alphabet $\mathcal{A}$,
- the possible belief sets $\mathbf{BS} = 2^{\mathcal{A}}$ contains all subsets of atoms, and
- $\mathbf{ACC}(P)$ is the set of $P$'s answer sets.

We now define a multi-context system according to Brewka and Eiter.

**Definition 2.10.** *A multi-context system $M = (C_1, \ldots, C_n)$ consists of a collection of contexts $C_i = (L_i, kb_i, br_i)$, where $L_i = (\mathbf{KB}_i, \mathbf{BS}_i, \mathbf{ACC}_i)$ is a logic, $kb_i \in \mathbf{KB}_i$ is a knowledge base, and $br_i$ is a set of $L_i$ bridge rules $r$ of the form*

$$a \leftarrow (c_1 : b_1), \ldots, (c_j : b_j), \sim(c_{j+1} : b_{j+1}), \ldots, \sim(c_m : b_m) \;, \tag{2.11}$$

*where $1 \leq c_k \leq n$, $b_k$ is an element of some belief set of $L_{c_k}$, $1 \leq k \leq m$, and $kb \cup \{a\} \in \mathbf{KB}_i$ for each $kb \in \mathbf{KB}_i$.*

We call a *context atom* $(c_k : b_k)$ or its default negation $\sim(c_k : b_k)$ a *context literal*. Let the atom $head(r) = a$ be the head of $r$ and the set $body(r) = \{(c_1 : b_1), \ldots, (c_j : b_j), \sim(c_{j+1} : b_{j+1}), \ldots, \sim(c_m : b_m)\}$ the *body* of $r$. For a set of context literals $S$, define $S^+ = \{(c : b) \mid (c : b) \in S\}$, $S^- = \{(c : b) \mid \sim(c : b) \in S\}$, and $S|_c = \{b \mid (c : b) \in S\}$. The set of atoms occurring in a set of bridge rules $br_i$, i.e., $\bigcup_{r \in br_i} \{b \mid (c : b) \in body(r)^+ \cup body(r)^-\} \cup \{head(r)\}$ is denoted by $atom(br_i)$.

Intuitively, context literals in bridge rules refer to information of another contexts. Bridge rules can thus modify the knowledge base, depending on what is believed or disbelieved in other contexts.

The semantics of an MCS is given by its equilibria, viz., acceptable belief sets, one from each context, which respect all bridge rules. More formally, for an MCS $M = (C_1, \ldots, C_n)$ define a *belief state* $S = (S_1, \ldots, S_n)$ of $M$ such that each $S_i \in \mathbf{BS}_i$. A bridge rule $r$ of the form (2.11) is *applicable* in a belief state $S$ iff $body(r)^+|_{c_k} \subseteq S_{c_k}$ and $body(r)^-|_{c_k} \cap S_{c_k} = \emptyset$ for all $1 \leq k \leq m$.

**Definition 2.11.** *A belief state* $S = (S_1, \ldots, S_n)$ *of an MCS* $M = (C_1, \ldots, C_n)$ *is an* equilibrium *iff for* $1 \leq i \leq n$:

$$S_i \in \mathbf{ACC}_i(kb_i \cup \{head(r) \mid r \in br_i,\ r \text{ is applicable in } S\}).$$

For multi-context systems with ASP-logics, we will assume that the underlying ASP logics $L_i$ are defined over pairwise distinct alphabets $\mathcal{A}_i$. A major difference to traditional answer set programming is that bridge-rules allow for a certain form of self-justification of atoms.

**Example 2.6.** *Consider an MCS* $M = (C_1, C_2)$ *with ASP logics over alphabets* $\mathcal{A}_1 = \{a\}$, $\mathcal{A}_2 = \{b\}$. *Suppose* $kb_1 = kb_2 = \emptyset$, $br_1 = \{a \leftarrow (2 : b)\}$, *and* $br_2 = \{b \leftarrow (1 : a)\}$. *One can check that* $(\{a\}, \{b\})$ *and* $(\emptyset, \emptyset)$ *are equilibria of* $M$, *while the answer set program* $P = \{a \leftarrow b, b \leftarrow a\}$ *has just a single answer set, that is* $\emptyset$.

Dao-Tran et al.'s distributed algorithms for nonmonotonic MCS $M = (C_1, \ldots, C_n)$ computes partial equilibria w.r.t. a context $C_k$, i.e., parts of potential equilibria of the system which contain coherent information from all contexts in the import closure of $C_k$. In the following, we recall some basic notions of Dao-Tran et al. (2010). The *import neighbourhood* of a context $C_k$ is the set

$$In(k) = \{c \mid (c : b) \in body(r), r \in br_i\}$$

The *import closure* $IC(k)$ is defined as the smallest set $S$ such that

- $k \in S$,

- $i \in S$ implies $In(i) \subseteq S$.

Let $\epsilon \notin \bigcup_{i=1}^{n} \mathbf{BS}_i$ be a new symbol representing the value 'unknown'. A *partial belief state* of $M$ is a sequence $S = (S_1, \ldots, S_n)$, such that $S_i \in \mathbf{BS}_i \cup \{\epsilon\}$, for $1 \leq i \leq n$.

**Definition 2.12.** *A partial belief state is a* partial equilibrium *of* $M$ *w.r.t. a context* $C_k$ *iff for* $1 \leq i \leq n$:

$$\begin{array}{ll} S_i \in \mathbf{ACC}_i(kb_i \cup \{head(r) \mid r \in br_i,\ r \text{ is applicable in } S\}) & \text{if } i \in IC(k), \\ S_i = \epsilon & \text{if } i \notin IC(k). \end{array}$$

Clearly, a partial equilibrium of $M$ w.r.t. context $C_k$ does not necessarily extend to an equilibrium of $M$. This is the case, for instance, if $\mathbf{BS}_i = \emptyset$ for some $i \notin IC(k)$. However, a partial equilibrium of $M$ w.r.t. $C_k$ is an equilibrium of the subsystem $M(k)$ defined by $IC(k)$.

For combining partial belief states $S = (S_1, \ldots, S_n)$ and $T = (T_1, \ldots, T_n)$ of $M$, their *join* is defined as the partial belief state $(U_1, \ldots, U_n)$ with

$$\begin{array}{ll} U_i = S_i & \text{if } T_i = \epsilon \vee T_i = S_i, \\ U_i = T_i & \text{if } T_i \neq \epsilon \wedge S_i = \epsilon. \end{array}$$

Roughly, one can compute partial equilibria of $M = (C_1, \ldots, C_n)$ w.r.t. context $C_k$ as follows:

1. Starting from context $C_k$, Dao-Tran et al.'s algorithm visits the import closure of $C_k$ by expanding the import neighbourhood at each context in a depth-first search with loop-detection.

2. A leave context returns partial belief states to the invoking context.

3. Intermediate contexts consistently combine their beliefs with partial belief states returned from their neighbours.

This algorithm can potentially be extended to equilibria of $M$. For a detailed description of the algorithm, we refer to (Dao-Tran et al., 2010).

# 3 Group Theoretic Background

Symmetries are studied in terms of groups. A *group* is an abstract algebraic structure $(G, *)$, where $G$ is a set closed under a binary associative operation $*$ such that there is a *unit* element and every element has a unique *inverse*.

**Definition 3.1.** *A group $(G, *)$ is a set $G$ with a binary operation $* : G \times G \to G$ that have the following three properties:*

- *the operation $*$ is* associative*, i.e. $\forall a, b, c \in G : (a * b) * c = a * (b * c)$,*

- *there is a* unit *element $e \in G$ such that $\forall a \in G : a * e = e * a = a$,*

- *for every $a \in G$ there exists a unique* inverse *$a^{-1} \in G$ such that $a * a^{-1} = a^{-1} * a = e$.*

*A* subgroup *is a subset of a group that is closed under the group operation, and is therefore a group itself.*

Often, we abuse notation and refer to the group $G$, rather than to the structure $(G, *)$. A compact representation of a group is given through generators.

**Definition 3.2.** *A set of group elements such that any other group element can be expressed in terms of their product is called a* generating set *or set of generators, and its elements are called generators. A generator is* redundant *if it can be expressed in terms of other generators. An* irredundant *generating set, by definition, does not contain redundant generators.*

An irredundant set of generators provides an extremely compact representation of a group. In fact, representing a group by a generating set always ensures exponential compression.

**Theorem 3.1** (Lagrange, from Elementary Group Theory; Hall, 1959)**.** *The size of any subgroup of any finite group $G$ must divide the size of $G$.*

We denote the size of a group $G$ as $|G|$.

**Corollary 3.2** (Aloul et al., 2002)**.** *Any irredundant generating set for a finite group $G$, such that $|G| > 1$, contains at most $\log_2 |G|$ elements.*

To relate different groups, we recall some more notion from algebra.

**Definition 3.3.** *A mapping $f : G \to H$ between to groups $(G, *)$ and $(H, \circ)$ is a* homomorphism *iff for and $a, b \in G$ we have $f(a * b) = f(a) \circ f(b)$. A homomorphism for which an inverse exists that is also a homomorphism, is called an* isomorphism. *If an isomorphism exists, the two groups $G$ and $H$ are called* isomorphic. *An isomorphism of a group with itself is called an* automorphism.

Since we can describe groups in terms of generators, it is important to know that isomorphisms preserve generators.

**Theorem 3.3** (Aloul et al., 2002). *Any group isomorphism maps sets of generators to sets of generators, and maps irredundant sets of generators to irredundant sets of generators.*

In our context, the most important group is the group of permutations.

**Definition 3.4.** *A* permutation *of a set $S$ is a bijection $\pi : S \to S$.*

Indeed, the set of permutations form a group under composition, denoted as $\Pi(S)$. It is easy to see that the composition of two permutations is a permutation, that the composition of permutations is associative, that the composition with the *identity* never changes a permutation, and that every permutation has a unique inverse. The image of $a \in S$ under a permutation $\pi$ is denoted as $a^\pi$, and for a set $X \subseteq S$ define $X^\pi = \{a^\pi \mid a \in X\}$.

**Definition 3.5.** *The* orbit *of $a \in S$ under a permutation $\pi \in \Pi(S)$ are the set of elements of $S$ to which $a$ can be mapped by (repeatedly) applying $\pi$.*

Note that orbits define an equivalence relation on elements in $S$. Analogously, for vectors $v = (v_1, v_2, \ldots, v_k) \in S^k$ define $v^\pi = (v_1^\pi, v_2^\pi, \ldots, v_k^\pi)$, for sets $X = \{a_1, a_2, \ldots, a_k\} \subseteq S$ define $X^\pi = \{a_1^\pi, a_2^\pi, \ldots, a_k^\pi\}$, and for sets of sets $X = \{X_1, X_2, \ldots, X_k\}$ such that $X_i \subseteq S$ for $1 \leq i \leq k$ define $X^\pi = \{X_1^\pi, X_2^\pi, \ldots, X_k^\pi\}$. Permutations can be expressed in tabular form, for example,

$$\pi = \left( \begin{array}{cccc} a_1 & a_2 & \ldots & a_n \\ a_1^\pi & a_2^\pi & \ldots & a_n^\pi \end{array} \right)$$

denotes a permutation that maps $a_1$ to $a_1^\pi$, etc. More often, we will make use of the *cycle notation* where a permutation is a product of disjoint cycles. A cycle $(a_1\ a_2\ a_3\ \ldots\ a_n)$ means that the permutation maps $a_1$ to $a_2$, $a_2$ to $a_3$, and so on, finally $a_n$ back to $a_1$. An element that does not appear in any cycle is understood as being mapped to itself. Finally, we define the *support* of a permutation (McKay, 1981) as those elements that are not mapped to themselves.

**Graph Automorphism Problems**

In graph theory, the symmetries are studied in terms of graph automorphism. We consider directed graphs $G = (V, E)$, where $V$ is a set of vertices and $E \subseteq V \times V$ is a set of directed edges. Intuitively, an automorphism of $G$ is a permutation of its vertices that maps edges to edges, and non-edges to non-edges, preserving edge orientation. More formally, we define as follows.

**Definition 3.6.** *An* automorphism *or a* symmetry *of a graph* $G = (V, E)$ *is a permutation* $\pi \in \Pi(V)$ *such that* $(u, v) \in E$ *iff* $(u, v)^\pi \in E$.

A further extension considers vertex colourings, where symmetries must map each vertex into a vertex with the same colour.

**Definition 3.7.** *Given a partition of the vertices* $\rho(V) = \{V_1, V_2, \ldots, V_k\}$. *An* automorphism *or a* symmetry *of a coloured graph $G$ is a symmetry $\pi$ of $G$ such that* $\rho(V)^\pi = \rho(V)$.

We will think of the partition $\rho$ as a *colouring* of the vertices.

**Example 3.1.** *Consider the graph* $G = (\{u, v, w\}, \{(u, v), (u, w), (v, u), (v, w)\})$ *with partition* $\{\{u, v\}, \{w\}\}$. *G's only nontrivial symmetry is* $\pi = (u\ v)$.

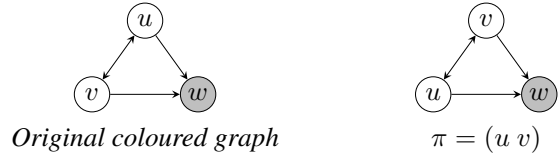

*Original coloured graph*      $\pi = (u\ v)$

The *(coloured) graph automorphism* problem is to find all symmetries of a given graph, for instance, in terms of generators. It is not known to have any polynomial time solution, and is conjectured to be strictly between the complexity classes P and NP (Babai, 1995), thus potentially easier than computing answer sets. Practical algorithms for computing graph automorphism groups have been implemented in the systems NAUTY (McKay, 1981), SAUCY (Darga et al., 2004; 2008), and BLISS (Junttila and Kaski, 2007).

**Symmetry in Constraint Satisfaction Problems**

Intuitively, a symmetry of a CSP is a transformation of its components that leaves the CSP unchanged. A common type of symmetry are value symmetries, which just act on values.

**Definition 3.8.** *A* value symmetry *is a bijection on values of constraint variables that preserve solutions.*

For example, the colours in a *graph colouring* problem can be freely permuted in any solution. If such a symmetry acts globally on values, we call it a *global value symmetry*.

**Symmetry in Answer Set Programming**

By a symmetry of an answer set program we mean a permutation of its atoms that does not change the logic program, in particular, maps rules to rules. In principle, such a permutation can affect arbitrarily many atoms at once, for instance, as in the case of a complete cyclic shift. For a rule $r$ of the form (2.1) and a permutation $\pi$ define

$$r^\pi = a_1^\pi; \ldots; a_\ell^\pi \leftarrow b_1^\pi, \ldots, b_m^\pi, \sim c_1^\pi, \ldots, \sim c_n^\pi$$

For a set of rules $P$, i.e., a logic program, define $P^\pi = \{r^\pi \mid r \in P\}$.

**Definition 3.9.** *A symmetry of a logic program $P$ is a permutation $\pi \in \Pi(atom(P))$ such that $\pi(P) = P$.*

By definition, a symmetry of a logic program preserves answer sets.

**Example 3.2.** *Reconsider $P_1$ from Example 2.1, and $\pi = (a\ b)$. Since $\pi(P_1) = P_1$, $\pi$ is a symmetry of $P_1$.*

**Example 3.3.** *There are four symmetries in the* all-interval series *problem: (1) the identity, (2) reversing the series (variable symmetry), (3) reflecting the series by subtracting each element from $n - 1$ (value symmetry), and (4) doing both. It is easy to see that (2) and (3) form a group of generators. Indeed, we can find both symmetries in our encoding (see Example 2.2) given in cycle notation below.*

$$\begin{aligned}
\pi_2 =\ & (v_{1,0}\ v_{n,0})\ (v_{1,1}\ v_{n,1})\ \ldots\ (v_{1,n-1}\ v_{n,n-1}) \\
& \ldots \\
& (v_{\lfloor n/2 \rfloor,0}\ v_{\lceil n/2 \rceil,0})\ (v_{\lfloor n/2 \rfloor,1}\ v_{\lceil n/2 \rceil,1})\ \ldots\ (v_{\lfloor n/2 \rfloor,n-1}\ v_{\lceil n/2 \rceil,n-1}) \\
& (d_{1,1}\ d_{n-1,1})\ (d_{1,2}\ d_{n-1,2})\ \ldots\ (d_{1,n-1}\ d_{n-1,n-1}) \\
& \ldots \\
& (d_{\lfloor (n-1)/2 \rfloor,1}\ d_{\lceil (n-1)/2 \rceil,1})\ (d_{\lfloor (n-1)/2 \rfloor,2}\ d_{\lceil (n-1)/2 \rceil,2}) \\
& \ldots\ (d_{\lfloor (n-1)/2 \rfloor,n-1}\ d_{\lceil (n-1)/2 \rceil,n-1}) \\
\pi_3 =\ & (v_{1,0}\ v_{1,n-1})\ (v_{1,1}\ v_{1,n-2})\ \ldots\ (v_{n,\lfloor (n-1)/2 \rfloor}\ v_{n,\lceil (n-1)/2 \rceil}) \\
& \ldots \\
& (v_{n,0}\ v_{n,n-1})\ (v_{n,1}\ v_{n,n-2})\ \ldots\ (v_{n,\lfloor (n-1)/2 \rfloor}\ v_{n,\lceil (n-1)/2 \rceil})
\end{aligned}$$

*Intuitively, the cycles in the first three lines of $\pi_2$ simply swap the first and the last variable, the second and the last but one variable, etc., value by value to reverse the series, where the remaining cycles adjust the auxiliary variables, i.e., swap the differences value by value, respectively. The cycles in $\pi_3$ swap the values $0$ and $n - 1$, $1$ and $n - 2$, etc., for each variable to reflect the series. Obviously, the permutations $\pi_2$ and $\pi_3$ represent (2) and (3), respectively, and do not change the logic program.*

### Symmetry in Multi-Context Systems with ASP Logics

We will extend our notion of a symmetry to multi-context systems instantiated with ASP logics. For bridge rules $r$ of the form (2.11) define

$$r^\pi = a^\pi \leftarrow (c_1 : b_1^\pi), \ldots, (c_j : b_j^\pi), \sim(c_{j+1} : b_{j+1}^\pi), \ldots, \sim(c_m : b_m^\pi).$$

and for a set of bridge rules $br$ define $br^\pi = \{r^\pi \mid r \in br\}$.

**Definition 3.10.** *Let $M = (C_1, \ldots, C_n)$ be an MCS such that all $L_i$ are ASP logics over $\mathcal{A}_i$. A symmetry of $M$ is a permutation $\pi \in \Pi(\bigcup_{i=1}^n \mathcal{A}_i)$ such that $\pi(kb_i) = kb_i$ and $\pi(br_i) = br_i$, for $1 \leq i \leq n$.*

**Example 3.4.** *Consider an MCS $M = (C_1, C_2)$ with ASP logics over alphabets $\mathcal{A}_1 = \{a, b\}$, $\mathcal{A}_2 = \{c, d\}$. Suppose $kb_1 = kb_2 = \emptyset$, and*

$$br_1 = \left\{ \begin{array}{l} a \leftarrow \sim(2:c) \\ b \leftarrow \sim(2:d) \end{array} \right\}, \qquad br_2 = \left\{ \begin{array}{l} c \leftarrow \sim(1:a) \\ d \leftarrow \sim(1:b) \end{array} \right\}.$$

*One can check that $(\{a, b\}, \emptyset)$, $(\{a\}, \{d\})$, $(\{b\}, \{c\})$, and $(\emptyset, \{c, d\})$ are equilibria. Observe, that a symmetry of $M$ is given through $\pi = (a\ b)\ (c\ d)$. In particular, the equilibria $(\{a\}, \{d\})$, $(\{b\}, \{c\})$ are symmetric.*

**Example 3.5.** *The MCS from Example 2.6 has only one symmetry: the identity.*

Sometimes, a symmetry affects only atoms from the belief set of a single context, i.e., behaves like the identity for the atoms of all other contexts. We call such a symmetry a local symmetry.

**Definition 3.11.** *Let $M = (C_1, \ldots, C_n)$ be an MCS such that all $L_i$ are ASP logics over $\mathcal{A}_i$. A symmetry $\pi$ of $M$ is* local *for context $C_i$ iff $\pi(a) = a$ for all $a \in dom(\pi) \setminus \mathcal{A}_i$.*

As a special case, the identity is local for all contexts of $M$.

**Example 3.6.** *Consider an MCS $M = (C_1, C_2)$ with ASP logics over alphabets $\mathcal{A}_1 = \{a, b\}$, $\mathcal{A}_2 = \{c, d, e, f\}$. Suppose*

$$kb_1 = \emptyset, \qquad\qquad br_1 = \left\{ \begin{array}{l} a \leftarrow \sim(2:c) \\ b \leftarrow \sim(2:d) \end{array} \right\},$$

$$kb_2 = \left\{ \begin{array}{l} e \leftarrow c, \sim e \\ f \leftarrow c, \sim f \end{array} \right\}, \qquad br_2 = \left\{ \begin{array}{l} c \leftarrow \sim(1:a) \\ d \leftarrow \sim(1:b) \end{array} \right\}.$$

*One can check that $(\{a\}, \{d\})$, $(\{b\}, \{c, e\})$ and $(\{b\}, \{c, f\})$ are equilibria. Observe, that a symmetry of $M$ is given through $\pi = (e\ f)$, which is also local for $C_2$. In particular, the equilibria $(\{b\}, \{c, e\})$ and $(\{b\}, \{c, f\})$ are symmetric.*

Similar to equilibria, we define the notion of partial symmetries, which are parts of potential symmetries of the system.

**Definition 3.12.** *Let $M = (C_1, \ldots, C_n)$ be an MCS such that all $L_i$ are ASP logics over $\mathcal{A}_i$. A permutation $\pi$ of the elements in $S \subseteq \bigcup_{i=1}^{n} \mathcal{A}_i$ is a* partial symmetry *of $M$ w.r.t. the set of contexts $C = \{C_{i_1}, \ldots, C_{i_m}\}$ iff the following conditions hold, for $1 \leq k \leq m$:*

- *$\mathcal{A}_{i_k} \cup atom(br_{i_k}) \subseteq S$,*
- *$\pi(kb_{i_k}) = kb_{i_k}$, and*
- *$\pi(br_{i_k}) = br_{i_k}$.*

For combining partial symmetries $\pi$ and $\sigma$, we define their *join* $\pi \bowtie \sigma$ as the partial symmetry $\theta$, if $\pi(a) = \sigma(a)$ for all $a \in dom(\pi) \cap dom(\sigma)$, and where

$$\theta(a) = \begin{cases} \pi(a) & \text{if } a \in dom(\pi), \\ \sigma(a) & \text{if } a \in dom(\sigma). \end{cases}$$

Note that, $\pi \bowtie \sigma$ is void, i.e. undefined, if $\pi$ and $\sigma$ behave different for some $a \in dom(\pi) \cap dom(\sigma)$. The *join* of two sets of partial symmetries is then naturally defined as $\Pi \bowtie \Sigma = \{\pi \bowtie \sigma \mid \pi \in \Pi, \sigma \in \Sigma\}$.

**Example 3.7.** *Consider an MCS $M = (C_1, C_2)$ with ASP logics over alphabets $\mathcal{A}_1 = \{a, b, c, d\}$, $\mathcal{A}_2 = \{e, f\}$. Suppose that*

$$kb_1 = \left\{ \begin{array}{l} a \leftarrow \sim b \\ b \leftarrow \sim a \\ c \leftarrow \sim d \\ d \leftarrow \sim c \end{array} \right\} \qquad br_1 = \emptyset,$$

$$kb_2 = \emptyset, \qquad br_2 = \left\{ \begin{array}{l} e \leftarrow (1 : a) \\ f \leftarrow (1 : b) \end{array} \right\}.$$

*One can check the following:*

- $\pi_1 = \begin{pmatrix} a & b & c & d \\ b & a & d & c \end{pmatrix}$ *is a partial symmetry of $M$ w.r.t. $\{C_1\}$,*

- $\pi_2 = \begin{pmatrix} a & b & e & f \\ b & a & f & e \end{pmatrix}$ *is a partial symmetry of $M$ w.r.t. $\{C_2\}$,*

- $dom(\pi_1) \cap dom(\pi_2) = \{a, b\}$, *and* $\pi_1(a) = \pi_2(a)$, $\pi_1(b) = \pi_2(b)$,

- $\pi_3 = \pi_1 \bowtie \pi_2 = \begin{pmatrix} a & b & c & d & e & f \\ b & a & d & c & f & e \end{pmatrix}$ *is a partial symmetry of $M$ w.r.t. $\{C_1, C_2\}$.*

*Furthermore, $\pi_3$ is a symmetry of $M$.*

# 4 Symmetry Detection

Our approach for detecting symmetries of a logic program is through reduction to, and solution of, an associated *graph automorphism* problem. Our techniques are based on the *body-atom graph* $(V, E_0 \cup E_1, E_2)$ of a logic program $P$, that is, a directed graph with vertices $V = body(P) \cup atom(P)$, and labelled edges $E_0 = \{(\beta, a) \mid a \in atom(P), \beta \in body(a)\}$, $E_1 = \{(a, \beta) \mid \beta \in body(P), a \in \beta^+\}$, and $E_2 = \{(a, \beta) \mid \beta \in body(P), a \in \beta^-\}$. The body-atom graph has been shown to be a suitable representation of a logic program (Linke, 2003). However, we modify the body-atom graph by introducing additional vertices for negated atoms to circumvent labelled edges, and construct a 3-coloured graph as follows:

1. In our graph encoding every atom in $atom(P)$ is represented by two vertices of colour 1 and 2 that correspond to the positive and negative literals, respectively.

2. Every rule is represented by a body vertex of colour 3, a set of directed edges that connect the vertices of the literals that appear in the rule's body to its body vertex, and a set of directed edges that connect the body vertex to the vertices of the atoms (positive literals) that appear in the head of the rule.

3. To ensure consistency, that is, $a$ maps to $b$ if and only if $\sim a$ maps to $\sim b$ for any atoms $a$ and $b$, vertices of opposite literals are mated by a directed edge from the positive literal to the negative literal.

The choice of three vertex colours insures that body vertices can only be mapped to body vertices, and positive (negative) literal vertices can only be mapped to positive (negative) lit-eral nodes. To conclude, given a logic program $P$ consisting of $m$ bodies and $l$ literals over $n$ atoms, the graph encoding for detecting symmetries of $P$ is constructed by $m + 2n$ vertices and $l + n$ edges. Figure 4.1 illustrates the general structure of a rule $r$ of the form (2.1) as a body-atom-graph (left), where $\beta$ is the body vertex. Straight lines represent edges in $E_0 \cup E_1$, curly lines represent edges in $E_2$. On the right is the general structure of a 3-coloured graph construction of $r$. Vertices of colour 1, 2, and 3 are represented by empty circles, filled circles, and empty squares, respectively. Figure 4.2 provides an example.

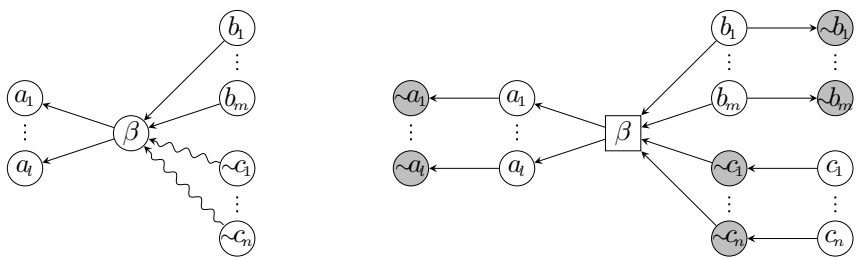

Figure 4.1: General structure of a rule as a body-atom-graph and its 3-coloured graph encoding.

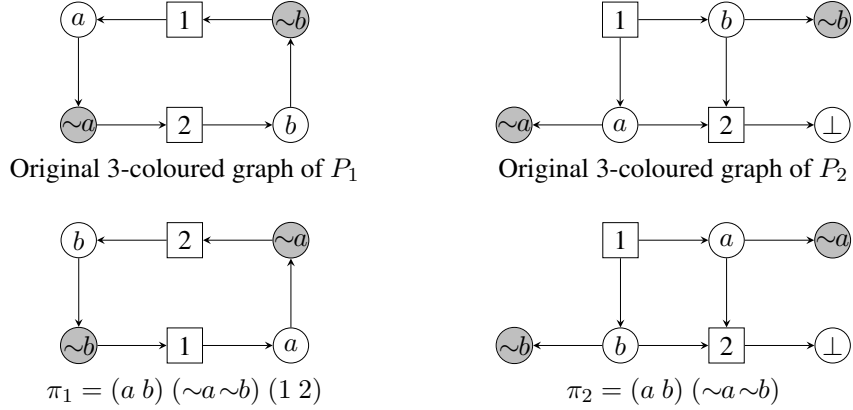

Figure 4.2: 3-coloured graph constructions and resulting symmetries for the example logic programs $P_1$ and $P_2$.

**Theorem 4.1.** *The symmetries of a logic program correspond one-to-one to the symmetries of its 3-coloured graph encoding.*

*Proof.* ($\Rightarrow$) We begin by showing that any symmetry of a logic program corresponds to a symmetry of the constructed 3-coloured graph. Such a graph symmetry will map vertices of the same colour and edges to edges. In particular, if $a$ maps to $b$, then $\sim a$ maps to $\sim b$, and the edge $(a, \sim a)$ maps to the edge $(b, \sim b)$. Since $a$ and $b$, and $\sim a$ and $\sim b$, have the same colour, the symmetry is preserved. The same can be said about the other edges between vertices of different colours: In a logic program, $a$ and $b$ might also be connected with one or more body vertices. These connections would also be swapped at the respective vertices. Again, only vertices of the same colour are mapped one to another. Thus, a consistent mapping of atoms in the program, when carried over to the graph, must preserve the colours of the vertices.

($\Leftarrow$) We now show that every symmetry in the graph corresponds to a symmetry of the logic program. It is not hard to see because we use one colour for positive literals, one for negative literals, and one for bodies. Hence, a graph symmetry must map (1) positive literal vertices to other such, and negative literal vertices to negative literal vertices, and body vertices to body vertices, and (2) the body edges of a vertex to body edges of its mate. This is consistent with symmetries of the logic program mapping atoms to atoms, and bodies to bodies, i.e., rules to rules. To prove Boolean consistency, i.e., if $a$ maps to $b$ then $\sim a$ maps to $\sim b$, we recall that every edge from a vertex of colour 2 to a vertex of colour 1 is a Boolean consistency edge of the form $(a, \sim a)$. Since every such edge can only map to another such edge, a mapping $a$ to $b$ leaves no choice for $(a, \sim a)$ but to map to $(b, \sim b)$ because $(b, \sim b)$ is the only edge that connects $b$ to another vertex of the same colour as $\sim a$. □

**Theorem 4.2.** *The symmetry groups of the logic program and its 3-coloured graph encoding are isomorphic.*

*Proof Sketch.* The proof consists of the straightforward verification that the one-to-one mapping constructed in the proof of Theorem 4.1 is a homomorphism. □

**Corollary 4.3.** *Sets of symmetry generators of the 3-coloured graph encoding correspond one-to-one to sets of symmetry generators of the logic program.*

*Proof.* By Theorem 3.3 and Theorem 4.2. □

Since GAP algorithms are sensitive to the number of vertices of an input graph, our construction can be optimised to reduce the number of graph vertices while preserving its symmetries. A first simplification is achieved by modelling rules with an empty body and a single head atom, so-called *facts*, by a (forth) colour for the vertex corresponding to the head atom instead of using (empty) body vertices. Furthermore, rules with a single head atom and a 1-literal body are modelled using a directed edge from the vertex corresponding to the literal of the body to the vertex corresponding to the head atom. Observe that this optimisation may connect a literal vertex to a positive literal vertex. Still, unintended mappings between 1-literal body edges and consistency edges remain impossible, since consistency edges connect positive literal vertices to their negative mates. For the special case of a 1-literal body and an empty head, we connect the literal vertex to the special node '$\perp$'.

We extend our graph encoding to integrity constraints, choice rules and cardinality constraints. No changes are necessary to cover integrity constraints. Also, the structure of a choice rule is encoded like a rule, i.e, is represented by a body vertex, a set of directed edges that connect the vertices of the literals that appear in the choice rule's body to its body vertex, and a set of directed edges that connect the body vertex to the vertices of the literals that appear in the head of the rule. To distinguish choice rules from rules a new colour 5 is introduced for their body vertices. An extension to cardinality constraints of the form (2.4) has to consider the bound $k$. Hence, we colour its body vertex by $k + 5$ to ensure that the literals of two cardinality constraints can be mated only if their bound is equal. Furthermore, each cardinality constraints is represented by a set of directed edges that connect the vertices of the literals $b_1, \ldots, b_m, \sim c_1, \ldots, \sim c_n$, that appear in its body, to its body vertex. Figure 4.3 (left) illustrates the general structure of a coloured graph construction of a choice rule of the form (2.3). Vertices of colour 1, 2, and 5 are represented by empty circles, filled circles, and filled squares, respectively. On the right is the general structure of a coloured graph construction of a cardinality constraint of the form (2.4). Vertices of colour 1, 2, and $k + 5$ are represented by empty circles, filled circles, and empty diamonds, respectively.

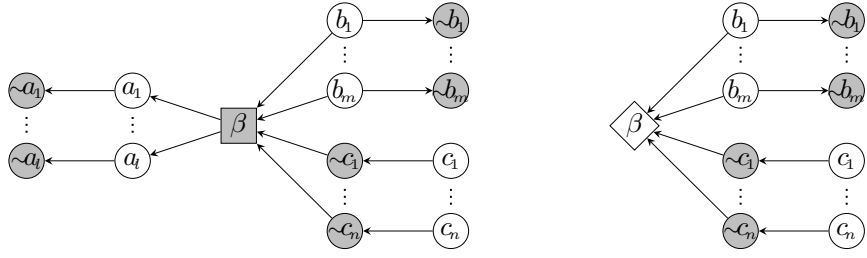

Figure 4.3: The general structure of a coloured graph construction of a choice rule and a cardinality constraint.

**Symmetry Detection for Multi-Context Systems with ASP Logics**

In the following, we provide a distributed algorithm for detecting symmetries of an MCS $M = (C_1, \ldots, C_n)$ with ASP logics. Our methods are inductive, i.e., we first take a local stance starting by detecting partial symmetries of $M$ w.r.t. $\{C_i\}$ for $1 \leq i \leq n$, viz., each context alone, and then successively try to combine partial symmetries.

For symmetry detection in a context $C_i$, we extend our approach for traditional answer set programs, i.e., we encode symmetry detection as a graph automorphism problem where the coloured graph is constructed by previously defined steps 1-3, and 4-6 defined as follows:

4. Every context atom $(c : b)$ that occurs in $br_i$ is represented by two vertices of a new colour $c$ and $c+1$ that correspond to the positive and negative context literals, respectively. This ensures that context atoms map to context atoms of the same context only.

5. Bridge rules are represented as traditional ASP rules, i.e., by a body vertex of a colour 4, a set of directed edges that connect the vertices of the literals that appear in the rule's body to its body vertex, and a set of directed edges that connect the body vertex to the vertices of the atoms (positive literals) that appear in the head of the rule.

6. To ensure consistency, that is again, $(c : b_i)$ maps to $(c : b_j)$ if and only if $\sim(c : b_i)$ maps to $\sim(c : b_j)$ for any context atoms $(c : b_i)$ and $(c : b_j)$, vertices of opposite literals are mated by a directed edge from the positive literal to the negative literal.

**Example 4.1.** *One can check that the symmetries of the graph shown in Figure 4.4 are the identity and $(a\ b)\ (c\ d)\ (1\ 2)$. Therefore, the partial symmetries of the MSC $M$ from Example 3.6 w.r.t. $\{C_1\}$, are the identity and $\pi_{1,1} = (a\ b)\ (c\ d)$.*

**Example 4.2.** *One can check that the symmetries of the graph shown in Figure 4.5 are the identity and $(e\ f)\ (1\ 2)$. Therefore, the partial symmetries of the MSC $M$ from Example 3.6 w.r.t. $\{C_2\}$, are the identity and $\pi_{1,1} = (e\ f)$.*

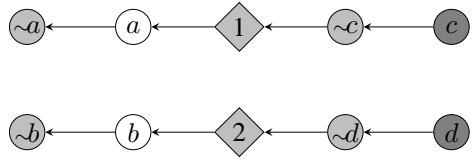

Figure 4.4: Coloured graph construction of context $C_1$ from the MSC in Example 3.6.

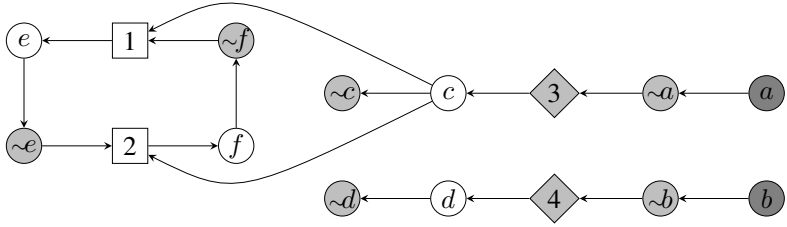

Figure 4.5: Coloured graph construction of context $C_2$ from the MSC in Example 3.6.

**Theorem 4.4.** *Let $M = (C_1, \ldots, C_n)$ be an MCS such that all $L_i$ are ASP logics over $\mathcal{A}_i$. The partial symmetries of $M$ w.r.t. $\{C_k\}$ correspond one-to-one to the symmetries of $C_k$'s coloured graph encoding.*

*Proof.* The proof is symmetric to the one of Theorem 4.1. Therefore we only provide arguments regarding bridge rules and context atoms. ($\Rightarrow$) A partial symmetries of $M$ w.r.t. $\{C_k\}$ will map context atoms to context atoms of the same context. Since they have the same colour, the symmetry is preserved for corresponding vertices and consistency edges. The same can be said about body vertices and edges representing bridge rules, since the body vertices have incoming edges from context literal vertices with their respective colour only, and vertices of the same colour are mapped one to another. Thus, a consistent mapping of atoms in $C_k$, when carried over to the graph, must preserve symmetry.

($\Leftarrow$) We now show that every symmetry in the graph corresponds to a partial symmetries of $M$ w.r.t. $\{C_k\}$. Recall that we use one colour for positive context literals from each context, one for negative context literals from each context, and one for bodies. Hence, a graph symmetry must map (1) positive context literal vertices to other such from the same context, and negative literal vertices to negative literal vertices from the same context, and body vertices to body vertices, and (2) the body edges of a vertex to body edges of its mate. This is consistent with partial symmetries of $M$ w.r.t. $\{C_k\}$ mapping context atoms to context atoms, and bodies to bodies, i.e., bridge rules to bridge rules. □

In the remainder of this section, we define a distributed algorithm for computing symmetries of an MCS. We follow Dao-Tran et al. (2010) by taking a local stance, i.e., we consider a context $C_k$ and those parts of the system that are in the import closure of $C_k$ to compute (potential) symmetries of the system. To this end, we design an algorithm whose instances run independently at each context node and communicate with other instances for exchanging sets of partial symmetries. This provides a method for distributed symmetry building.

**Algorithm**: DSD($H$) at context $C_k = (L_k, kb_k, br_k)$ with ASP logic $L_i$
**Input**: Visited contexts $H$.
**Data**: Cache $c(k)$.
**Output**: The set of accumulated partial symmetries $\Pi$.

> **if** $c(k)$ is not initialised **then**
>     $c(k) \leftarrow \text{GAP}(C_k)$;
> $H \leftarrow H \cup \{k\}$;
> $\Pi \leftarrow c(k)$;
> **foreach** $i \in In(k) \setminus H$ **do**
>     $\Pi \leftarrow \Pi \bowtie C_i.\text{DSD}(H)$;
> **return** $\Pi$;

Algorithm 4.1: The distributed symmetry detection algorithm.

The idea is as follows: starting from a context $C_k$, we visit the import closure of $C_k$ by expanding the import neighbourhood at each context, maintaining the set of visited contexts in a set $H$, the *history*, until a leaf context is reached, or a cycle is detected by noticing the presence of a neighbour context in $H$. A leaf context $C_i$ simply computes all partial symmetries of $M$ w.r.t. $\{C_i\}$ by encoding symmetry detection as a graph automorphism problem, and invoking a dedicated GAP solver. Then, it returns the results to its parent (invoking context), for instance, in form of permutation cycles. The results of intermediate contexts $C_i$ are partial symmetries of $M$ w.r.t. $\{C_i\}$, which can be joined, i.e., consistently combined, with partial symmetries from their neighbours, and resulting in partial symmetries of $M$ w.r.t. $\{C_j \mid j \in IC(i)\} \cup \{C_i\}$. In particular, the starting context $C_k$ returns its partial symmetries joined with the results from its neighbours, as a final result.

We assume that each context $C_k$ has a background process, e.g. a daemon, that waits for incoming requests with history $H$, upon which it starts the computation outlined in our algorithm shown in Figure 4.1. We write $C_i.\text{DSD}(H)$ to specify that we send $H$ to the process at context $C_i$ and wait for its return message. This process also serves the purpose of keeping the cache $c(k)$ persistent. We use the primitive $\text{GAP}(C_k)$ which computes partial symmetries of $M$ w.r.t. $\{C_k\}$ via reduction to a graph automorphism problem, as described before. Our algorithm proceeds in the following way:

1. Check the cache for partial symmetries of $M$ w.r.t. $\{C_k\}$.

2. If imports from neighbour contexts are needed, then request partial symmetries from all neighbours and join them (previously visited contexts excluded). This can be performed in parallel. Also, partial symmetries can be joined in the order neighbouring contexts do answer.

3. Return partial symmetries of $M$ w.r.t. $\{C_i \mid i \in IC(k)\}$.

Correctness of our approach is provided by the following result.

**Theorem 4.5.** *Let $M = (C_1, \ldots, C_n)$ be an MCS such that all $L_i$ are ASP logics over $\mathcal{A}_i$. Then, $\pi \in C_k.\text{DSD}(\emptyset)$ iff $\pi$ is a partial symmetry of $M$ w.r.t. $\{C_k\} \cup \{C_i \mid i \in IC(k)\}$.*

*Proof.* We start by showing that the combination of partial symmetries of $M$ are partial symmetries of $M$.

**Lemma 4.6.** *Let $M = (C_1, \ldots, C_n)$ be an MCS such that all $L_i$ are ASP logics over $\mathcal{A}_i$. Let $\Pi_i$ the set of partial symmetries of $M$ w.r.t. $C_i$. The partial symmetries of $M$ w.r.t. $\{C_{i_1}, \ldots, C_{i_m}\}$ are given through $\Pi_{i_1} \bowtie \ldots \bowtie \Pi_{i_m}$. In particular, a partial symmetry of $M$ w.r.t. $\{C_1, \ldots, C_n\}$ is a symmetry of $M$.*

*Proof of Lemma 4.6.* We prove by induction on $m$. Base case: $\Pi_{i_1}$ are the partial symmetries of $M$ w.r.t. $\{C_1\}$, by Theorem 4.4. Induction step: Assume the partial symmetries of $M$ w.r.t. $\{C_{i_1}, \ldots, C_{i_m}\}$ are $\Pi = \Pi_{i_1} \bowtie \ldots \bowtie \Pi_{i_m}$, and the set of partial symmetries of $M$ w.r.t. $\{C_{i_{m+1}}\}$ is $\Pi_{i_{m+1}}$. ($\Rightarrow$) Let $\pi \in \Pi$ and $\sigma \in \Pi_{i_{m+1}}$ such that $\pi(a) = \sigma(a)$ for all $a \in dom(\pi) \cap dom(\sigma)$. Then, by Definition 3.12, $\theta = \pi \bowtie \sigma$ is a partial symmetry of $M$ w.r.t. $\{C_{i_1}, \ldots, C_{i_m}, C_{i_{m+1}}\}$. To be more precise, $\theta(kb_{i_j}) = kb_{i_j}$ and $\theta(br_{i_j}) = br_{i_j}$ since $\pi(kb_{i_j}) = kb_{i_j}$ and $\pi(br_{i_j}) = br_{i_j}$ for all $1 \leq j \leq m$, and $\theta(kb_{i_{m+1}}) = kb_{i_{m+1}}$ and $\theta(br_{i_{m+1}}) = br_{i_{m+1}}$ since $\sigma(kb_{i_{m+1}}) = kb_{i_{m+1}}$ and $\sigma(br_{i_{m+1}}) = br_{i_{m+1}}$.

($\Leftarrow$) Let $\theta$ be any partial symmetry of $M$ w.r.t. $\{C_{i_1}, \ldots, C_{i_m}, C_{i_{m+1}}\}$. Trivially, by Definition 3.12, $\theta$ is also a partial symmetry of $M$ w.r.t. $\{C_{i_1}, \ldots, C_{i_m}\}$, and $M$ w.r.t. $\{C_{i_{m+1}}\}$. Hence, $\theta \in \Pi$ and $\theta \in \Pi_{i_{m+1}}$, and $\theta \in \Pi \bowtie \Pi_{i_{m+1}}$. In conclusion, $\Pi \bowtie \Pi_{i_{m+1}}$ are the partial symmetries of $M$ w.r.t. $\{C_{i_1}, \ldots, C_{i_m}, C_{i_{m+1}}\}$.

Finally, let $\pi$ be a partial symmetry of $M$ w.r.t. $\{C_1, \ldots, C_n\}$. By Definition 3.12 we have $dom(\pi) \subseteq \bigcup_{i=1}^{n} \mathcal{A}_i$ (upper bound for the domain of partial symmetries), and $\mathcal{A}_i \subseteq dom(\pi)$ (lower bound for domain of partial symmetries), for $1 \leq i \leq n$. Hence, $\pi$ is a permutation of exactly the elements in $\bigcup_{i=1}^{n} \mathcal{A}_i$. Given this, and since $\pi(kb_i) = kb_i$ and $\pi(br_i) = br_i$ holds for $1 \leq i \leq n$, i.e., all contexts in $M$, $\pi$ is a symmetry of $M$. □

We can now prove Theorem 4.5. ($\Rightarrow$) We prove soundness, i.e., if $\pi \in C_k.\text{DSD}(\emptyset)$ then $\pi$ is a partial symmetry of $M$ w.r.t. $\{C_k\} \cup \{C_i \mid i \in IC(k)\}$. We show by structural induction on the topology of an MCS, and start with acyclic MCS $M$. Base case: $C_k$ is a leaf with $br_k = \emptyset$ and $In(k) = \emptyset$. By Theorem 4.4, we compute all partial symmetries of $M$ w.r.t. $\{C_k\}$, i.e., $c(k) \leftarrow \text{GAP}(C_k)$ in Algorithm 4.1. Induction step: Assume the import neighbourhood of context $C_k$ is $In(k) = \{i_1, \ldots, i_m\}$ and

$$\begin{aligned} \Pi_k &= \text{GAP}(C_k) \\ \Pi_{i_1} &= C_{i_1}.\text{DSD}(H \cup \{k\}) \\ &\vdots \\ \Pi_{i_m} &= C_{i_m}.\text{DSD}(H \cup \{k\}) \end{aligned}$$

By Lemma 4.6, the partial symmetries of $M$ w.r.t. $\{C_k\} \cup \{C_i \mid i \in IC(k)\}$ are $\Pi = \Pi_k \bowtie \Pi_{i_1} \bowtie \ldots \bowtie \Pi_{i_m}$, as computed by $\Pi \leftarrow \Pi \bowtie C_i.\text{DSD}(H)$ in the loop of Algorithm 4.1.

The proof for cyclic $M$ works similarly. In a run we eventually end up in a context $C_i$ such that $i \in H$ again. In that case, calling $C_i.\text{DSD}(H)$ is discarded, which breaks the cycle.

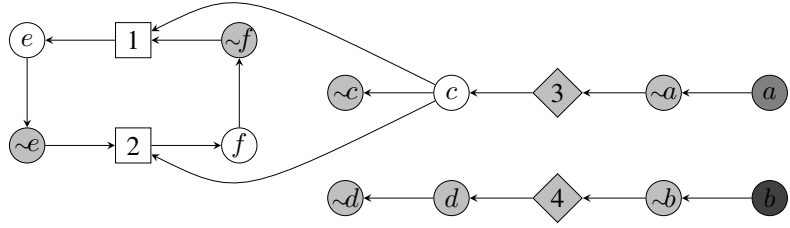

Figure 4.6: Modified coloured graph construction of context $C_2$ from the MSC in Example 3.6.

However, partial symmetries not including $C_i$ are propagated through the system to the calling context $C_i$ which combines the intermediate results with partial symmetries of $M$ w.r.t. $\{C_i\}$.

($\Leftarrow$) We give now a proof sketch for completeness. Let $\pi$ be a partial symmetry of $M$ w.r.t. $\{C_k\} \cup \{C_i \mid i \in IC(k)\}$. We show $\pi \in C_k.\mathrm{DSD}(\emptyset)$. The proof idea is as follows: we proceed as in the soundness part by structural induction on the topology of $M$, and in the base case for a leaf context $C_k$, by Theorem 4.4, we get that $\mathrm{GAP}(C_k)$ gives us all partial symmetries of $M$ w.r.t. $\{C_k\}$. By definition of a symmetry, $\pi|_{\mathcal{A}_k \cup atom(br_k)}$ is a partial symmetry of $M$ w.r.t. $\{C_k\}$. Symmetric arguments hold for partial symmetries of $M$ w.r.t. $\{C_k\} \cup \{C_i \mid i \in IC(k)\}$. $\square$

**Example 4.3.** *Reconsider $M$ from Example 3.6. Suppose the user invokes $C_1.\mathrm{DSD}(\emptyset)$ to trigger the symmetry detection process. When called the first time, the process of context $C_2$ determines partial symmetries $\Pi_1$ of $M$ w.r.t. $\{C_2\}$, that are given by identity and $\pi_1 = (a\ b)\ (c\ d)$ (see Figure 4.4), and triggers its neighbours (only context $C_2$). Eventually, the process at $C_2$ returns partial symmetries $\Pi_2$ of $M$ w.r.t. $\{C_i \mid i \in IC(i)\} \cup \{C_2\}$ ($= \{C_2\}$), that are the identity and $\pi_2 = (e\ f)$ (see Figure 4.5). Finally, the process at $C_1$ computes the join $\Pi_1 \bowtie \Pi_2$, that consists of the identity and $(e\ f)$. These results are returned to the user.*

To compute local symmetries only, we further modify our approach by assigning a unique colour to each context atom and each atom that is referenced in other contexts, i.e., context atoms cannot be mapped.

**Example 4.4.** *One can check that the symmetries of the graph shown in Figure 4.6 are the identity and $(e\ f)\ (1\ 2)$. Therefore, the local symmetries of the MSC $M$ from Example 3.6 in $C_1$, are the identity and $\pi_2 = (e\ f)$. Hence, $\pi_2$ is a symmetry of $M$.*

**Theorem 4.7.** *Let $M = (C_1, \ldots, C_n)$ be an MCS such that all $L_i$ are ASP logics over $\mathcal{A}_i$. The local symmetries of $M$ in $C_k$ correspond one-to-one to the symmetries of $C_k$'s modified coloured graph encoding.*

*Proof.* The proof is symmetric to the one of Theorem 4.1. Therefore we only provide arguments regarding bridge rules and context atoms.

($\Rightarrow$) A local symmetry of $M$ in $C_k$ behaves on context atoms like the identity mapping. Since they have unique colours, this property is preserved for corresponding vertices and consistency edges. The same can be said about body vertices and edges representing bridge rules, since the body vertices have incoming edges from context literal vertices with their respective colour only, and only vertices of the same colour are mapped one to another. Thus, a consistent mapping of atoms in $C_k$, when carried over to the graph, must preserve symmetry.

($\Leftarrow$) We now show that every symmetry in the graph corresponds to a local symmetry of $M$ in $C_k$. Recall that we use a unique colour for each positive context literal. Hence, a graph symmetry behaves like the identity mapping on context literal vertices and body vertices which represent bridge rules. This naturally carries over to the body edges of a context literal vertex. Extend the symmetry to all atoms in $\bigcup_{i=1}^{n} \mathcal{A}_i$ such that previously undefined mappings behave like the identity mapping. This is consistent with local symmetries of $M$ in $C_k$. □

# 5 Symmetry Breaking

## 5.1 Symmetry-breaking Constraints

Recall that a symmetry $\pi$ of a logic program $P$ defines equivalence classes on the atoms in $P$ (orbits). This naturally extends to Boolean assignments, where, for signed literals $\mathbf{T}a$ and $\mathbf{F}a$, we define $(\mathbf{T}a)^\pi = \mathbf{T}(a^\pi)$ and $(\mathbf{F}a)^\pi = \mathbf{F}(a^\pi)$. Hence, symmetries induces equivalence classes in the solution space of a problem: Given an answer set of $P$, all sets to which it can be mapped by symmetries, must be answer sets of $P$. Similarly, symmetries always map non-answer sets to non-answer sets. Therefore, it is sufficient to reason about one representative from every equivalence class.

Symmetry breaking amounts to selecting some representatives from every equivalence class and constructing rules, composed into a symmetry-breaking constraint, that is only satisfied on those representatives. A *full* SBC selects exactly one representative from each orbit, otherwise we call an SBC *partial*. The most common approach is to order all elements from the solution space lexicographically, and to select the lexicographically smallest element, the *lex-leader*, from each orbit as its representative (c.f. Crawford et al., 1996; Aloul et al., 2002; 2003a;b). A *lex-leader symmetry-breaking constraint* (LL-SBC) is an SBC that is satisfied only on the lex-leaders of orbits.

We will assume a total ordering on the atoms $a_1, a_2, \ldots, a_n$ of a logic program's alphabet $\mathcal{A}$ and consider the induced lexicographic ordering on the truth assignments, i.e., their interpretation as unsigned integers. The construction of a lex-leader SBC is accomplished by encoding a *permutation constraint* (PC) for every permutation $\pi$, where

$$\text{PC}(\pi) = \bigwedge_{1 \leq i \leq n} \left[ \bigwedge_{1 \leq j \leq i-1} (a_j = a_j{}^\pi) \right] \rightarrow (a_i \leq a_i{}^\pi).$$

The lex-leader symmetry-breaking constraint that breaks every symmetry in a logic program can now be constructed by conjoining all of its permutation constraints.

$$\text{LL-SBC}(\Pi) = \bigwedge_{\pi \in \Pi} \text{PC}(\pi)$$

Through *chaining*, which includes additional atoms, we achieve a PC representation that is linear in the number of atoms (Aloul et al., 2003a):

$$
\begin{aligned}
\text{PC}(\pi) &= (a_1 \leq a_1^\pi) \wedge c_{\pi,2} \\
c_{\pi,i} &\equiv (a_{i-1} \geq a_{i-1}^\pi) \rightarrow (a_i \leq a_i^\pi) \wedge c_{\pi,i+1} \qquad 1 < i \leq n \\
c_{\pi,n+1} &\equiv \top
\end{aligned}
$$

**Theorem 5.1** (Crawford et al., 1996). *For a group of symmetries $\Pi$, the truth assignments that satisfy* LL-SBC($\Pi$) *are the lexicographically smallest representatives from each class of truth assignments that can be mapped to each others by elements from $\Pi$.*

Finally, we encode above permutation constraint in ASP, denoted $P(\pi)$, that is satisfied for the lex-leader of the orbit induced by $\pi$ as follows:

$$
\begin{aligned}
&\leftarrow a_1, \sim a_1^\pi && (5.1) \\
&\leftarrow c_{\pi,2} && (5.2) \\
c_{\pi,i} &\leftarrow a_{i-1}, a_i, \sim a_i^\pi && (5.3) \\
c_{\pi,i} &\leftarrow \sim a_{i-1}^\pi, a_i, \sim a_i^\pi && (5.4) \\
c_{\pi,i} &\leftarrow a_{i-1}, c_{\pi,i+1} && (5.5) \\
c_{\pi,i} &\leftarrow \sim a_{i-1}^\pi, c_{\pi,i+1} && (5.6) \\
c_{\pi,n+1} &\leftarrow && (5.7)
\end{aligned}
$$

where $1 < i \leq n$. Note that new atoms are introduced, thus extending the alphabet of $P$. Correctness is provided by the following theorem.

**Theorem 5.2.** *Let $\pi$ be a symmetry over atoms $\mathcal{A}$. For any truth assignment $\mathbf{A}$, $\mathbf{A}$ satisfies* $\text{PC}(\pi)$ *iff* $\forall \delta \in \Delta_{P(\pi)} \cup \Theta_{P(\pi)} : \delta \not\subseteq \mathbf{A}$.

*Proof Sketch.* Using the one-to-one correspondance between nogoods and clauses (Bayardo and Schrag, 1997), the proof consists of verifying that the nogoods represented by rules (5.3–5.6) correspond to the clausal form of $c_{\pi,i}$ and, even more obvious, the nogoods represented by rules (5.1–5.2), and (5.7) correspond to the clausal form of $\text{PC}(\pi)$, $\top$ respectively. □

A careful analysis reveals some possibilities to reduce the size of permutation constraints. The first corresponds to atoms that are mapped to themselves by the permutation, i.e., $a_i^\pi = a_i$. This makes the consequent of the implication unconditionally true. For sparse symmetries, one can significantly reduce the size of the permutation constraint with a restriction of the PC construction to only those atoms that are in the support of $\pi$. Second, also for atoms $a$ and $b$ such that both appear in $P$ as facts, and $a^\pi = b$, the consequent $a \leq a^\pi$ is satisfied.

A third possibility corresponds to the lexicographically largest atom in each cycle of $\pi$. Assume a cycle $(a_s \ldots a_e)$ on the atoms of some index set $\{a, \ldots, e\}$. Using equality propagation on the portion of the permutation constraint where $i = e$, we get $(a_s = a_e) \rightarrow (a_e \leq a_s)$ which is tautologous. Hence, we can further restrict the index set in the PC by excluding the lexicographically largest atom in each cycle.

**Example 5.1.** *We illustrate our PC encoding on the symmetries detected for the previous examples $P_1$ and $P_2$. Since both permutations $\pi_1$ and $\pi_2$ (Figure 4.2) map $a$ to $b$ and vice versa, they share the same LL-SBC which is as simple as follows, assuming $a$ is lexicographically smaller than $b$:*

$$\leftarrow a, \sim b$$

*Observe that the ordering on the atoms of a logic program $P$ induces a preference relation on the answer sets of $P$ under symmetry breaking. Here, the ordering selects $\{b\}$ as the representative of the set of all answer sets symmetric to $\{b\}$, hence, eliminating the answer set $\{a\}$.*

**Partial Symmetry Breaking**

Breaking all symmetries may not speed up search because there are often exponentially many of them. A better trade-off may be provided by breaking enough symmetries (Crawford et al., 1996). We explore partial SBCs, i.e., we do not require that SBCs are satisfied by lex-leading assignments only (but we still require that all lex-leaders satisfy SBCs). Irredundant generators are good candidates because they cannot be expressed in terms of each other, and implicitly represent all symmetries. Hence, breaking all symmetries in a generating set can eliminate all problem symmetries. However, this does not hold in general, e.g., different generating sets of the group of a logic program's symmetries may lead to different pruning (Katsirelos et al., 2009).

**Example 5.2.** *Consider a logic program $P$ with interchangeable atoms $a_1, a_2, a_3, a_4$, for instance*

$$\{a_1, a_2, a_3, a_4\} \leftarrow$$
$$\leftarrow a_1, a_2, a_3, a_4$$

*An irredundant generating set for $\Pi(P)$ is the pair swap $(a_1\ a_2)$ and the rotation $(a_1\ a_2\ a_3\ a_4)$. To break the symmetry $(a_1\ a_2)$ we post the permutation constraint*

$$\leftarrow a_1, \sim a_2$$

*To break the symmetry $(a_1\ a_2\ a_3\ a_4)$ we post*

$$\begin{array}{lll}
\leftarrow a_1, \sim a_2 & c_0 \leftarrow a_1, a_2, \sim a_3 & c_1 \leftarrow a_2, a_3, \sim a_4 \\
\leftarrow c_0 & c_0 \leftarrow a_1, c_1 & c_1 \leftarrow a_2, c_2 \\
c_2 \leftarrow & c_0 \leftarrow \sim a_2, c_1 & c_1 \leftarrow \sim a_3, c_2
\end{array}$$

*However, these two permutation constraints do not eliminate all symmetries. For instance, they permit both answer sets $\{a_2, a_4\}$ and its symmetry $\{a_3, a_4\}$. There is an alternative irredundant generating set which breaks all symmetries, that is $\{(a_1\ a_2), (a_2\ a_3), (a_3\ a_4)\}$. We can break these three symmetries with*

$$\leftarrow a_1, \sim a_2$$
$$\leftarrow a_2, \sim a_3$$
$$\leftarrow a_3, \sim a_4$$

*eliminating all symmetries of $P$.*

We can further relax symmetry breaking to $k$ supports from each permutation (Aloul et al., 2003a). For $k \leq n$ and $a_i$ is a support of permutation $\pi$, we define the *partial* permutation constraint:

$$\leftarrow a_1, \sim a_1{}^\pi$$
$$\leftarrow c_{\pi,2}$$
$$c_{\pi,i} \leftarrow a_{i-1}, a_i, \sim a_i{}^\pi$$
$$c_{\pi,i} \leftarrow \sim a_{i-1}{}^\pi, a_i, \sim a_i{}^\pi$$
$$c_{\pi,i} \leftarrow a_{i-1}, c_{\pi,i+1}$$
$$c_{\pi,i} \leftarrow \sim a_{i-1}{}^\pi, c_{\pi,i+1}$$
$$c_{\pi,k+1} \leftarrow$$

By restricting the construction of permutation constraints this way, we further reduce the size of partial SBC.

**Example 5.3.** *Consider the* all-interval series *problem encoded from Example 2.2 and the generators $\pi_2$ and $\pi_3$ from Example 3.3. The symmetry-breaking constraint, where both permutation constraints are restricted to the second support, is given through the following, where $c_0, \ldots, c_3$ are new atoms.*

$$\leftarrow v_{1,0}, \sim v_{1,n-1} \qquad \qquad \leftarrow v_{1,0}, \sim v_{n,0}$$
$$\leftarrow c_0 \qquad \qquad \leftarrow c_2$$
$$c_0 \leftarrow v_{1,0}, v_{1,1}, \sim v_{1,n-2} \qquad c_2 \leftarrow v_{1,0}, v_{1,1}, \sim v_{n,1}$$
$$c_0 \leftarrow v_{1,1}, \sim v_{1,n-1}, \sim v_{1,n-2} \qquad c_2 \leftarrow v_{1,1}, \sim v_{n,0}, \sim v_{n,1}$$
$$c_0 \leftarrow v_{1,0}, c_1 \qquad \qquad c_2 \leftarrow v_{1,0}, c_3$$
$$c_0 \leftarrow c_1, \sim v_{1,n-1} \qquad \qquad c_2 \leftarrow c_3, \sim v_{n,0}$$
$$c_1 \leftarrow \qquad \qquad c_3 \leftarrow$$

Our techniques can be easily extended to constraint answer set programming using our translation based approach (Drescher and Walsh, 2010a;b), where a constraint logic program is decomposed into a logic program under answer set semantics. Then generic symmetry detection and symmetry breaking can be applied. However, it is often reasonable to assume that the symmetries for a problem are known, and can be modelled a-priori.

## 5.2 Breaking Value Symmetry

For particular symmetries, there are more efficient breaking methods. We show here how to deal with value symmetries. Recall, a *value symmetry* is a bijection on values of constraint variables that preserve solutions. In this context, a pair of values is called *interchangeable* if they can be swapped in any solution. One can break all symmetries between a pair of interchangeable values $(d_j, d_k)$ in the scope of the constraint variables $v_1, \ldots, v_n$ using the *value precedence* constraint (Law and Lee, 2004)

$$precedence([d_j, d_k], [v_1, \ldots, v_n]) \qquad (5.8)$$

which holds iff the smallest index of a variable that takes the value $d_i$ is smaller than the smallest index of a variable that takes the value $d_k$, or $d_k$ is not taken by any variable $v_1, \ldots, v_n$, i.e.,

$$min(\{i \mid v_i = d_j\} \cup \{n+1\}) < min(\{i \mid v_i = d_k\} \cup \{n+2\}) \,.$$

We follow Walsh (2006) and encode this constraint by introducing a sequence of atoms $b_i$ for $1 \leq i \leq n$ (5.9), where $\mathbf{T}b_i$ if $v_l = d_j$ for some $l < i$. Based on the direct encoding, the *value precedence* constraint prevents us from assigning $\mathbf{T}e(v_i, d_k)$, representing $v_i = d_k$, unless $\mathbf{T}b_i$. Hence, we post the following sequence of constraints, i.e., rules (5.10–5.13) that hold iff $v_i = d_j$ implies $\mathbf{T}b_i$ (5.10), $v_i \neq d_k$ implies $b_i = b_{i+1}$ (5.11–5.12), and $\mathbf{F}b_i$ implies $v_i \neq d_k$. We also set $\mathbf{F}b_1$ (5.14).

$$\{b_1, \ldots, b_n\} \leftarrow \tag{5.9}$$
$$\leftarrow e(v_i, d_j), \sim b_{i+1} \qquad 1 \leq i < n \tag{5.10}$$
$$\leftarrow \sim e(v_i, d_j), b_i, \sim b_{i+1} \qquad 1 \leq i < n \tag{5.11}$$
$$\leftarrow \sim e(v_i, d_j), \sim b_i, b_{i+1} \qquad 1 \leq i < n \tag{5.12}$$
$$\leftarrow e(v_i, d_k), \sim b_i \qquad 1 \leq i \leq n \tag{5.13}$$
$$\leftarrow b_1 \tag{5.14}$$

**Theorem 5.3.** *Unit-propagation on (2.6) and (5.9–5.13) enforces domain consistency on Law and Lee's* precedence *constraint in $\mathcal{O}(nd)$ down any branch of the search tree.*

*Proof.* The proof of domain consistency is provided by Walsh (2006). It remains to prove the runtime. For each of the $n$ variables, there are $\mathcal{O}(d)$ nogoods resulting from (2.6) that can be propagated $\mathcal{O}(1)$ times down any branch of the search tree. Each propagation requires $\mathcal{O}(1)$ time. Rule (2.6) therefore take $\mathcal{O}(nd)$ down any branch of the search to propagate. There are $\mathcal{O}(n)$ nogoods resulting from (5.9–5.14) that each take $\mathcal{O}(1)$ time to propagate down any branch of the search tree. Therefore, the total running time is given by $\mathcal{O}(nd) + \mathcal{O}(n) = \mathcal{O}(nd)$. □

Many problems, however, involve multiple interchangeable values, not just two. For instance, we assign colours to vertices in the *graph colouring* problem (Example 2.5), all values are interchangeable. To break such symmetry, Law and Lee propose the *global value precedence* constraint

$$precedence([d_1, \ldots, d_m], [v_1, \ldots, v_n]) . \tag{5.15}$$

In what follows, we consider $[d_1, \ldots, d_m] = \bigcup_{1 \leq i \leq n} D_i$. Then, the *global value precedence* constraint (5.15) holds iff for all $1 \leq i < j < m$

$$min(\{i \mid v_i = d_i\} \cup \{n+1\}) < min(\{i \mid v_i = d_j\} \cup \{n+2\}) .$$

To propagate this constraint, Law and Lee (2004) suggest decomposing it into pairwise *value precedence* constraints of the form (5.8), i.e., $precedence([d_j, d_k], [v_1, \ldots, v_n])$ for all $j < k$. However, Walsh proved that such a decomposition hinders GAC propagation (Walsh, 2006). We propose instead a simple ASP encoding of $precedence([d_1, \ldots, d_m], [v_1, \ldots, v_n])$, inspired by Bacchus' automaton-based CNF decomposition of the *regular* constraint (Bacchus, 2007). (Note that the *global value precedence* constraint is an instance of the *regular* constraint.) A regular language $\mathcal{L}$ has an associated deterministic finite automaton (DFA) $M$ that accepts a string iff that string is a member of $\mathcal{L}$. Each $M$ is defined by the quintuple $(Q, \Sigma, \delta, q_0, F)$, where $Q$ is a finite set of automaton states, $\Sigma$ is an input alphabet, $\delta$ is a transition function $Q \times \Sigma \to Q$, $q_0$ is the initial state, and $F$ is a set of accepting states (Homer and Selman, 2001). Each input symbol $s \in \Sigma$ causes $M$ to perform a transition from its current state $q$ to the new state $\delta(q, s)$, where $M$ starts off in the state $q_0$ and a string $S$ over the alphabet $\Sigma$. $S$ is said to be accepted by the DFA $M$ if $M$ is in an accepting state after processing $S$. We encode a deterministic finite automaton $M = (Q, \Sigma, \delta, d_0, F)$ that accepts the input string $S = [v_1, \ldots, v_n]$ iff $precedence([d_1, \ldots, d_m], [v_1, \ldots, v_n])$ is satisfied. $M$ is defined by $\Sigma = \{d_1, \ldots, d_m\}$, $Q = \Sigma \cup \{d_0\}$, $F = \Sigma$, and $\delta(A(v_i), d_j) = max(A(v_i), d_j)$ provided $v_i \leq d_{j+1}$ for $0 \leq j < m$. A state diagram is given in Figure 5.1. For each step $i$ of $M$'s processing $1 \leq i \leq n$, and each state $d_j \in Q$, an atom $state(v_i, d_j)$ is introduced. This atom is true if $M$ is in state $d_j$ when processing input symbol $v_i$. Since $d_0$ is the initial state, $state(v_1, d_0)$ is given as a fact 5.16. The transition function is encoded in rules (5.17–5.19). Consider $M$ is in state $d_j$ and reads symbol $v_i$, i.e., $state(v_i, d_j)$ is true. If $v_i = d_{j+1}$, i.e., $i$ is the smallest index of a variable that takes the value $d_{j+1}$, then $\delta(v_i, d_{j+1}) = v_i$ and $state(v_i, d_{j+1})$ becomes true (5.17). If $v_i \neq d_{j+1}$, i.e., $i$ is not the smallest index of a variable that takes the value $d_{j+1}$, then $M$ remains in the state $d_j$ (5.18). $M$ rejects if $v_i > d_j$ (5.19).

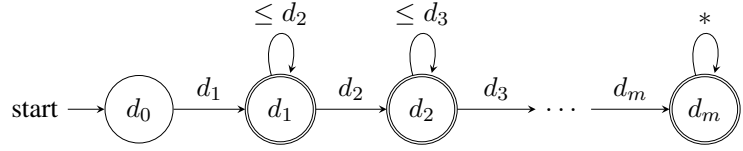

Figure 5.1: The state diagram of a DFA $M$ encoding the *global value precedence* constraint on multiple interchangeable values. A state $d_j$ represents the maximum value seen before reading the input symbol $v_i$. If $v_i$ takes the value $d_{j+1}$, $M$ goes in state $d_{j+1}$ and remains in $d_j$, otherwise, provided $v_i \leq d_{j+1}$. It is understood that all transitions not shown lead to a (rejecting) sink state.

$$state(v_1, d_0) \leftarrow \qquad (5.16)$$
$$state(v_{i+1}, d_{j+1}) \leftarrow state(v_i, d_j), e(v_i, d_{j+1}) \qquad 1 \leq i < n,\ 0 \leq j < m \qquad (5.17)$$
$$state(v_{i+1}, d_j) \leftarrow state(v_i, d_j), \sim e(v_i, d_{j+1}) \qquad 1 \leq i < n,\ 0 \leq j \leq m \qquad (5.18)$$
$$\leftarrow state(v_i, d_j), e(v_i, d_k) \qquad 1 \leq i \leq n,\ 1 \leq j < k+1 \leq m \qquad (5.19)$$

A logic program that contains an ASP encoding of $M$ has no answer set if $M$ rejects $S$, i.e., $precedence([d_1, \ldots, d_m], [v_1, \ldots, v_n])$ can not be satisfied. Furthermore, unit-propagation enforces domain consistency on the *global value precedence* constraint.

**Theorem 5.4.** *Unit-propagation on (2.6–2.8) and (5.16–5.19) enforces domain consistency on the* global value precedence *constraint in $\mathcal{O}(nmd)$ down any branch of the search tree.*

*Proof.* Suppose we have a set of domains for the constraint variables in which no pruning is possible and no domain is empty. First, if $v_i = d_k$ is not pruned by the *global value precedence* constraint it must have compatible values in the domains of the other variables, i.e., $v_1, \ldots, v_i, \ldots v_n$ is a sequence of inputs to the DFA $M$ representing $precedence([d_1, \ldots, d_m], [v_1, \ldots, v_n])$ which causes $M$ to transition through a sequence of states starting at the initial state $d_0$ and ending at an accepting state $d_m \in F$. Setting all of the state atoms of the form $state(v_j, d_\ell)$ and corresponding atoms of the form $e(v_j, d_\ell)$ to be true, and all other atoms to be false, we observe that all nogoods encoding rules (5.16–5.19) cannot be violated any more. Hence $\mathbf{T}e(v_i, d_k)$ is part of a satisfying truth assignment and since unit-propagation is sound it cannot force $\mathbf{F}e(v_i, d_k)$. We conclude by contraposition that if unit-propagation forces $\mathbf{F}e(v_i, d_k)$ then the *global value precedence* constraint prunes $v_i = d_k$ to achieve domain consistency.

Second, if no unit-propagation is possible and $\mathbf{F}e(v_i, d_k)$ has not been forced, then by rule (5.19) nogoods $\{\mathbf{T}state(v_i, d_j), \mathbf{T}e(v_i, d_k)\}$ are not unit, for all $j < k+1$. Hence the state variables $state(v_i, d_j)$, in particular the state variable $state(v_i, d_{k-1})$, also have not been falsified by unit-propagation. By (nogoods encoding) rules (5.17–5.18) state atoms $state(v_{i-1}, d_{k-1})$ and $state(v_{i+1}, d_{k+1})$ cannot be falsified. Hence, $e(v_{i-1}, d_{k-1}$ and $e(v_{i+1}, d_{k+1}$ cannot be falsified either. Continuing this way we arrive at an input sequence that includes $d_j \in dom(v_i)$ and that causes $M$ to transition from $d_0$ to an accepting state. That is, $v_i = d_j$ has compatible values in the domain of all the other variables, and the *global value precedence* constraint does not prune it.

Finally, if the *global value precedence* constraint has no satisfying assignment then no value $d \in dom(v_i)$ is compatible for any variable $v_i$. By above, unit-propagation will force $\mathbf{F}e(v_i, d)$ for every $d \in dom(v_i)$ thus violating rule (2.7) expressed in the nogood $\{\mathbf{F}v_i, d_1, \ldots, \mathbf{F}v_i, d_k\}$.

Now we address the runtime. For each of the $n$ variables, there are $\mathcal{O}(nd)$ nogoods resulting from (2.6) that take $\mathcal{O}(d)$ down any branch of the search to propagate. There are $\mathcal{O}(nmd)$ nogoods resulting from (5.16–5.19) that each take $\mathcal{O}(1)$ time to propagate down any branch of the search tree. Hence, the total runtime is given by $\mathcal{O}(nmd)$. By comparison, the total runtime of Law and Lee's decomposition into pairwise *value precedence* constraints is given by $\mathcal{O}(nm^2d)$. □

Since ASP solvers are sensitive to the size of rules in terms of literals, we want to optimize our encoding, and represent all values for $v_i$ allowed by $precedence([d_1, \ldots, d_m], [v_1, \ldots, v_n])$ instead of the value seen so far (expressed by the state $q_j$ of $M$ before reading $v_i$). Hence, for each pair $(v_i, d_j)$ we introduce a set of atoms $Q_{i,j} = \{allowed(v_i, d_j) \mid 1 \leq k \leq j+1\}$. Since $d_0$ is the initial state, the atom in $Q_{1,1} = \{allowed(v_1, d_1)\}$ is given as fact (5.20). Whenever a value $d_j$ is taken for the first time by $v_i$, rule (5.21) enforces $allowed(v_{i+1}, d_{j+1})$. In any case, values already seen are propagated through (5.22). Finally, $M$ rejects if $v_i$ takes a value which is not in the $allowed$-relation (5.23).

$$allowed(v_1, d_1) \leftarrow \qquad (5.20)$$
$$allowed(v_{i+1}, d_{j+1}) \leftarrow e(v_i, d_j) \qquad 1 \leq i < n,\ 1 \leq j < m \qquad (5.21)$$
$$allowed(v_{i+1}, d_j) \leftarrow allowed(v_i, d_j) \qquad 1 \leq i < n,\ 1 \leq j \leq m \qquad (5.22)$$
$$\leftarrow e(v_i, d_j), \sim allowed(v_i, d_j) \qquad 1 \leq i \leq n,\ 1 \leq j \leq m \qquad (5.23)$$

We refer to this ASP representation as our GAC encoding of the *global value precedence* constraint. Since domain consistency is the strongest type of local consistency, there can be no encoding that achieves more pruning. In particular, *global value precedence* prunes more values than symmetry-breaking in terms of generators (Katsirelos et al., 2009).

## 5.3 Distributed Symmetry-breaking Constraints

Symmetry breaking for distributed computation of (partial) equilibria of MCS requires distributed symmetry breaking techniques. Recall that a symmetry $\pi$ of an MCS $M$ potentially involves atoms from more than one context of $M$, except $\pi$ is local in a context $C_k$. To start with the general case, let $M = (C_1, \ldots, C_n)$ be an MCS such that all $L_i$ are ASP logics over $\mathcal{A}_i$, and $\pi$ be a symmetry of $M$. We will assume a total ordering on the atoms $a_1, a_2, \ldots, a_n$ in $\bigcup_{i=1}^{n} \mathcal{A}_i$ and consider the induced lexicographic ordering on the belief states. Based on our techniques for traditional answer set programming, we encode the *distributed permutation constraint* (distributed PC) that is satisfied for the lex-leading belief state induced by $\pi$ below. For a context $C_k$, we introduce intermediate atoms $b_{\pi,i-1}$ and $d_{\pi,i}$ to access information about $a_i \leq a_i^\pi$ and $c_{\pi,i+1}$ of another context. This might add further dependencies in form of bridge rules between previously independent contexts. In conclusion, for a context $C_k$, define the distributed permutation constraint $kb_k(\pi)$ and $br_k(\pi)$ as follows:

$$\left.\begin{array}{l} \leftarrow a_1, \sim a_1^\pi \\ \leftarrow c_{\pi,2} \end{array}\right\} \text{ in } kb_k(\pi), \text{ if } a_1 \in \mathcal{A}_k,$$

$$\left.\begin{array}{l} c_{\pi,i} \leftarrow a_{i-1}, a_i, \sim a_i^\pi \\ c_{\pi,i} \leftarrow \sim a_{i-1}^\pi, a_i, \sim a_i^\pi \\ c_{\pi,i} \leftarrow a_{i-1}, c_{\pi,i+1} \\ c_{\pi,i} \leftarrow \sim a_{i-1}^\pi, c_{\pi,i+1} \end{array}\right\} \text{ in } kb_k(\pi), \text{ if } a_{i-1}, a_i \in \mathcal{A}_k,$$

$$\left.\begin{array}{l} c_{\pi,i} \leftarrow a_{i-1}, b_{\pi,i-1} \\ c_{\pi,i} \leftarrow \sim a_{i-1}^\pi, b_{\pi,i-1} \\ c_{\pi,i} \leftarrow a_{i-1}, d_{\pi,i} \\ c_{\pi,i} \leftarrow \sim a_{i-1}^\pi, d_{\pi,i} \end{array}\right\} \text{ in } kb_k(\pi), \text{ if } a_{i-1} \in \mathcal{A}_k, a_i \in \mathcal{A}_{k+1},$$

$$\left.\begin{array}{l} b_{\pi,i-1} \leftarrow (k+1 : a_i), \sim(k+1 : a_i^\pi) \\ d_{\pi,i} \leftarrow (k+1 : c_{\pi,i+1}) \end{array}\right\} \text{ in } br_k(\pi), \text{ if } a_{i-1} \in \mathcal{A}_k, a_i \in \mathcal{A}_{k+1},$$

$$\left. c_{\pi,n+1} \leftarrow \phantom{} \right\} \text{ in } kb_k(\pi), \text{ if } a_n \in \mathcal{A}_k,$$

where $1 < i \leq n$. Observe that we have some options to eliminate tautologies (cf. Section 5.1). We now can define the lex-leader *distributed symmetry breaking constraint* (distributed SBC) by conjoining all distributed PC: Given a set of symmetries $\Pi$ of $M$, we construct a new MCS $M(\Pi) = (C_1(\Pi), \ldots, C_n(\Pi))$ over an extended alphabet, based on $M$, where $C_k(\Pi)$ extends $C_k$ by $kb_k(\Pi) = kb_k \cup \bigcup_{\pi \in \Pi} kb_k(\pi)$ and $br_k(\Pi) = br_k \cup \bigcup_{\pi \in \Pi} br_k(\pi)$. $M(\Pi)$ breaks all symmetries of $M$.

**Corollary 5.5.** *Let $\Pi$ be the symmetries of MCS $M = (C_1, \ldots, C_n)$ with all $L_i$ are ASP logics over $\mathcal{A}_i$. The belief states $S$ of $M(\Pi)$ are the lexicographically smallest representatives from each class of belief states of $M$ that can be mapped to each others by elements from $\Pi$.*

*Proof Sketch.* Verify that we achieve $P(\Pi)$ by replacing the intermediate atoms by the conditions they represent. Then Corollary 5.5 follows from Theorem 5.2. □

**Example 5.4.** *Reconsider the MCS $M$ and symmetry $\pi_3 = (a\ b)\ (c\ d)\ (e\ f)$ from Example 3.7. The equilibria of $M$ are $(\{a,c\},\{e\})$, $(\{a,d\},\{e\})$, $(\{b,c\},\{f\})$, and $(\{b,d\},\{f\})$. The MCS $M(\Pi)$, that breaks $\pi_3$ (among others), contains the following additional rules:*

$$
\left.\begin{aligned}
&\leftarrow a, \sim b \\
&\leftarrow c_{\pi_3,2} \\
c_{\pi_3,2} &\leftarrow c, b_{\pi_3,1} \\
c_{\pi_3,2} &\leftarrow \sim d, b_{\pi_3,1} \\
c_{\pi_3,2} &\leftarrow c, d_{\pi_3,2} \\
c_{\pi_3,2} &\leftarrow \sim d, d_{\pi_3,2}
\end{aligned}\right\} \text{ in } kb_1(\pi_3),
$$

$$
\left.\begin{aligned}
b_{\pi_3,1} &\leftarrow (2:e), \sim(2:f) \\
d_{\pi_3,2} &\leftarrow (2:c_{\pi_3,3})
\end{aligned}\right\} \text{ in } br_1(\pi_3), \text{ and}
$$

$$
\left.c_{\pi_3,3} \leftarrow \phantom{x}\right\} \text{ in } kb_2(\pi_3).
$$

*Observe that, because of the distributed PC defined above, $(\{a,c\},\{e\})$, $(\{a,d\},\{e\})$ are no equilibria of $M(\Pi)$. We can break the local symmetry $\pi_4 = (c\ d)$ of $M$ in $C_1$ by the distributed PC given through $kb_1(\pi_4) = \{\leftarrow c, \sim d\}$, $br_1(\pi_4) = kb_2(\pi_4) = br_2(\pi_4) = \emptyset$. The only equilibrium of $M(\Pi)$ is then $(\{b,d\},\{f\})$.*

Since in practice one is not interested in equilibria of the whole system $M$, but partial equilibria of $M$ w.r.t. $\{C_k\}$, i.e., equilibria of the subsystem $M(k)$, we suggest to detect and break symmetries of $M(k)$. Remark, partial symmetry breaking in terms of generators does not carry over to distributed SBCs.

**Theorem 5.6.** *The join of partial symmetries does not preserve generators.*

*Proof.* Consider an MCS $M$ with context $C_1$ such that the atoms $a, b, c, d$ can be freely permuted, and context $C_2$ such that the atoms $a$ and $c$ can be swapped. A generating set of partial symmetries of $M$ w.r.t. $\{C_1\}$ is given through $\Pi = \{(a\ b\ c\ d), (c\ d)\}$, and a generating set of $M$ w.r.t. $\{C_2\}$ is given through $\Sigma = \{(a\ d)\}$. By 4.6, $\Pi \bowtie \Sigma$ contain partial symmetries of $M$ w.r.t. $\{C_1, C_2\}$. Assume that $\Pi \bowtie \Sigma$ is a generating set of $M$ w.r.t $\{C_1, C_2\}$. Since $\Pi \bowtie \Sigma = \emptyset$, there are no partial symmetries of $M$ w.r.t. $\{C_1, C_2\}$. This contradicts to the observation that, for instance, the identity and $(a\ d)$ are partial symmetries of $M$ w.r.t. $\{C_1, C_2\}$. □

# 6 Evaluation

## 6.1 The SBASS System

Our approach to symmetry-breaking answer set solving has been implemented within the preprocessor SBASS, available at the (Potassco labs suite). The global architecture of SBASS is shown in Figure 6.1. It accepts a logic program $P$ in SMODELS format (Syrjänen) produced by a grounder, e.g. LPARSE, available at the (SMODELS suite), and GRINGO, available at the (Potassco suite). A first component, the Program Reader, takes care of creating an internal representation and encodes symmetry detection as a graph automorphism problem. Notably, the Program Reader also checks for duplicate edges in the graph encoding of $P$ which, otherwise, defect further processing. The actual search for an irredundant generating set of the group of symmetries of $P$, taking $P$'s graph encoding as input, is performed by the graph automorphism program SAUCY (2.1), available at the (SAUCY website), which is incorporated into SBASS. SAUCY sequentially returns graph symmetry generators as soon as they are detected. Each such symmetry is used to construct a PC, all of which result in an SBC. In turn, SBASS prints $P$ together with symmetry-breaking constraints, again in SMODELS format, which can be applied to any suitable answer set solver, e.g. SMODELS, available at the (SMODELS suite), and CLASP, available at the (Potassco suite). Note that SBASS provides several options, for instance, to print detected generators in cycle notation or statistics.

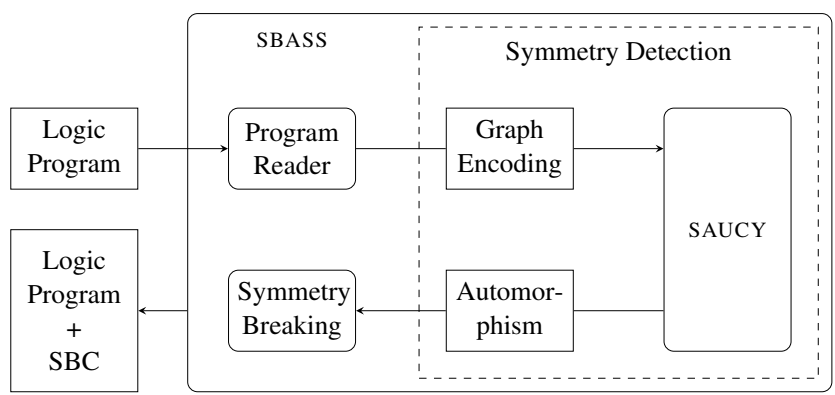

Figure 6.1: Global architecture of SBASS.

Table 6.1: Runtime results in seconds for *pigeon hole* problems using the disjunctive encoding.

| #$n$ | #gen. | SBASS | CLASP$_1^\pi$ | CLASP$_5^\pi$ | CLASP$^\pi$ | CLASP |
|---|---|---|---|---|---|---|
| 11 | 18 | 0.05 | 0.38 | 0.15 | 0.06 | 0.62 |
| 12 | 20 | 0.08 | 4.09 | 0.07 | 0.22 | 5.99 |
| 13 | 22 | 0.11 | 30.57 | 0.43 | 0.32 | 53.39 |
| 14 | 24 | 0.16 | 272.72 | 4.95 | 1.73 | 448.98 |
| 15 | 26 | 0.23 | — | 62.61 | 3.02 | — |
| 16 | 28 | 0.32 | — | — | 23.01 | — |
| 17 | 30 | 0.44 | — | — | 130.87 | — |

## 6.2 Experiments on Symmetry-breaking Answer Set Solving

To evaluate our approach, we conducted experiments on ASP encodings of several difficult combinatorial search problems. We use GRINGO (2.0.5) to generate our proposed encodings. Since our encodings are disjunctive, but tight, we make use of *shifting* (Gelfond et al., 1991) to provide an adequate encoding for the ASP solver CLASP, that are *normal* logic programs and its extensions. Experiments consider the answer set solver CLASP (1.3.2) on instances with symmetry breaking in terms of generators, i.e., instances preprocessed by SBASS, and without symmetry breaking. To explore the impact of partial PC, we restrict the construction of permutation constraints to $k$ supports per permutation, denoted as CLASP$_k^\pi$, using SBASS' option -size=k.

All tests were run on a 2.00 GHz PC under Linux, where each run was limited to 600 s time and 1 GB RAM, preprocessing excluded. However, we also report the runtime for SBASS and give the number of generators. The latter allows careful conclusions to be drawn with respect to the size of the search space implicitly pruned through symmetry breaking. In the following experiments we generally compare the runtime for testing the existence of an answer set to a given problem.

**Pigeon Hole Problems**

The *pigeon hole* problem is to show that it is impossible to put $n$ pigeons into $n-1$ holes if each pigeon must be put into a distinct hole. This problem is provably exponentially hard for any resolution based method (Urquhart, 1987), but is tractable using symmetries (all the pigeons are interchangeable and all the holes are interchangeable).

We chose a disjunctive encoding for the *pigeon hole* problem, where $p_{ij}$ is taken to mean that pigeon $i$ is assigned hole $j$ (Drescher et al., 2010):

$$p_{i,1}; p_{i,2}; \ldots; p_{i,n-1} \leftarrow \qquad i \in 1 \ldots n$$
$$\leftarrow p_{i,j}, p_{k,j} \qquad i < k$$

The runtimes for various sizes of $n$ are shown in Table 6.1. Although symmetry breaking has a positive impact, the runtime even with full PC is still exponentially growing with the number of pigeons. Here, symmetry breaking on the generating set returned by SAUCY does not break all problem symmetries. At this point, we should note that a given problem can be encoded in many *equivalent* logic programs (Lifschitz et al., 2001; Eiter and Fink, 2003), and with each different encoding our techniques may detect a different generating set. Therefore, we also tried an encoding of the *pigeon hole* problem based on the support encoding (Drescher and Walsh, 2010a;b):

$$
\begin{aligned}
\{p_{i,1}, p_{i,2}, \ldots, p_{i,n-1}\} &\leftarrow & i \in 1\ldots n \\
&\leftarrow \sim p_{i,1}, \sim p_{i,2}, \ldots, \sim p_{i,n-1} & i \in 1\ldots n \\
&\leftarrow 2\{p_{i,1}, p_{i,2}, \ldots, p_{i,n-1}\} & i \in 1\ldots n \\
&\leftarrow p_{i,j}, p_{k,j} & i < k
\end{aligned}
$$

Table 6.2: Runtime results in seconds for *pigeon hole* problems using the support encoding.

| #$n$ | #gen. | SBASS | CLASP$_1^\pi$ | CLASP$_5^\pi$ | CLASP$^\pi$ | CLASP |
|---|---|---|---|---|---|---|
| 11 | 19 | 0.03 | 8.70 | 0.02 | 0.02 | 47.28 |
| 12 | 21 | 0.04 | 66.57 | 0.03 | 0.03 | 397.01 |
| 13 | 23 | 0.07 | 540.26 | 0.09 | 0.03 | — |
| 14 | 25 | 0.08 | — | 0.77 | 0.04 | — |
| 15 | 27 | 0.12 | — | 5.91 | 0.05 | — |
| 16 | 29 | 0.17 | — | 47.98 | 0.06 | — |
| 17 | 31 | 0.22 | — | 520.39 | 0.13 | — |

This caused SAUCY to compute a different, obviously better set of generators, which consequently breaks all symmetry resulting in a polynomial runtime. (Observe the change in the number of generators.) As can be seen in Table 6.2, full PCs are essential to tackle the *pigeon hole* problem.

**Ramsey's Theorem**

*Ramsey's Theorem* states that for any pair of positive integers $(k, m)$ there exists a least positive integer $n$ such that, no matter how we colour the edges of the clique with $n$ vertices, $K_n$, using two colours, say blue and red, there is a sub-clique with $k$ vertices of colour blue or a sub-clique with $m$ nodes of colour red. Symmetries in *Ramsey's Theorem* are between the colours and the vertices in the sub-clique. *Ramsey's Theorem* is discussed in many articles (see, for instance, Graham and Rothschild, 1978) and can be found in the (Asparagus library) and the (CSP library).

We used the encoding by Leone et al. (2002), denoted as $R(k, m, n)$, to determine whether $n$ is not an integer for which the theorem holds. The problem $R(3, 5, n)$ is encoded as follows:

$$
\begin{aligned}
blue_{i,j}; red_{i,j} &\leftarrow & i, j \in 1\ldots n,\ i < j \\
&\leftarrow red_{i,j}, red_{i,k}, red_{j,k} & i, j, k \in 1\ldots n,\ i < j < k \\
&\leftarrow blue_{i,j}, blue_{i,k}, blue_{j,k}, \\
&\quad blue_{i,l}, blue_{j,l}, blue_{k,l}, \\
&\quad blue_{i,m}, blue_{j,m}, blue_{k,m}, blue_{l,m} & i, j, k, l, m \in 1\ldots n, \\
& & i < j < k < l < m
\end{aligned}
$$

Table 6.3: Average time for completed runs in seconds and the number of timeouts, if any, on *Ramsey's Theorem* instances, each shuffled 5 times. The *asterisk denotes instances that have no answer sets.

|  | #gen. | SBASS time | $CLASP_1^\pi$ time | $CLASP_1^\pi$ #t.out | $CLASP_5^\pi$ time | $CLASP_5^\pi$ #t.out | $CLASP^\pi$ time | $CLASP^\pi$ #t.out | CLASP time | CLASP #t.out |
|---|---|---|---|---|---|---|---|---|---|---|
| $R(3,5,13)$ | 11 | 0.06 | **0.01** |  | **0.01** |  | 0.03 |  | **0.01** |  |
| $R(3,5,14)^*$ | 12 | 0.10 | 3.58 |  | 1.23 |  | **0.49** |  | 354.25 |  |
| $R(3,6,17)$ | 15 | 1.18 | 0.12 |  | 0.12 |  | 0.14 |  | **0.11** |  |
| $R(3,6,18)^*$ | 16 | 1.87 | — | 5 | — | 5 | — | 5 | — | 5 |
| $R(4,4,17)$ | 15 | 0.26 | 0.73 |  | 0.12 |  | 0.50 |  | **0.07** |  |
| $R(4,4,18)^*$ | 16 | 0.37 | — | 5 | — | 5 | — | 5 | — | 5 |
| $R(4,5,23)$ | 21 | 5.43 | 4.23 |  | 2.29 |  | 2.05 |  | **1.32** |  |
| $R(4,5,24)$ | 22 | 7.15 | **77.64** |  | 208.66 | 1 | 180.96 | 3 | — | 5 |
| $R(4,5,25)^*$ | 23 | 9.54 | — | 5 | — | 5 | — | 5 | — | 5 |

Intuitively, the disjunctive rule guesses a colour for each edge. The first integrity constraint eliminates the colourings containing a red clique with 3 vertices, and the second integrity constraint eliminates the colourings containing a blue clique with 5 vertices.

In formerly hard cases, namely $R(3,5,14)$ and $R(4,5,24)$, symmetry breaking lead to significant pruning of the search space and yield solutions in a considerably short amount of time. The results presented in Table 6.3 suggest full PCs for unsatisfiable instances, but small, partial PCs for satisfiable instances.

### Graceful Graphs

A labelling $f$ of the vertices of a graph $(V, E)$ is *graceful* if $f$ assigns a unique label $f(v)$ from $\{0, 1, \ldots, |E|\}$ to each vertex $v \in V$ such that, when each edge $(v, w) \in E$ is assigned the label $|f(v) - f(w)|$, the resulting edge labels are distinct. The problem of determining the existence of a graceful labelling of a graph has been modelled as a CSP by Petrie and Smith (2003), and is an interesting application for symmetry-breaking answer set solving because the symmetries are different for each instance and cannot be modelled a-priori in general.

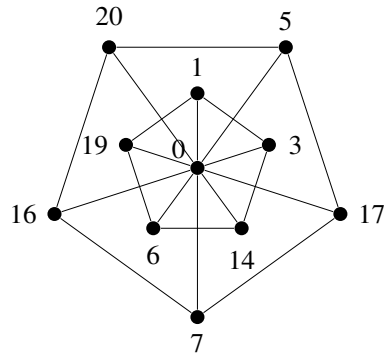

Figure 6.2: A graceful labelling of the double wheel graph $DW_5$.

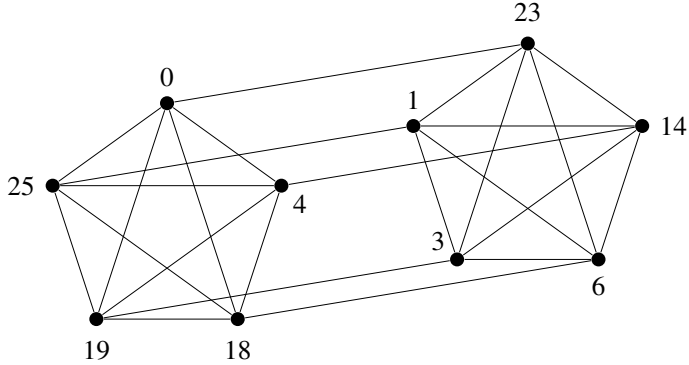

Figure 6.3: A graceful labelling of the graph $K_5 P_2$.

Our experiments consider graphs $DW_n$ and $K_n P_m$. The *double wheel* graph $DW_n$ is composed of two copies of a cycle with $n$ vertices, each connected to a central hub (Figure 6.2). The two wheels $W_n$, each have rotation and reflection symmetries. The labels of the two cycles can also be interchanged. The graph $K_n P_m$ is the cross-product of the clique $K_n$ and the path $P_m$ (Figure 6.3). It consists of $m$ copies of $K_n$, with corresponding vertices in the $m$ cliques also forming the vertices of a path $P_m$. Symmetries of the graph are simultaneous rotations of the cliques and inter-clique permutations.

As can be seen in Table 6.4, we achieve speed-up on the unsatisfiable instance $DW_3$. For the other instances, all of which are satisfiable, no complete traversal of the search space is necessary, and the branching heuristic used in our approach sometimes appears to be misled by the extra variables introduced in CLASP$_k^\pi$. That explains some of the variability in the runtimes. However, we still observe a substantial impact of our symmetry breaking techniques on the difficult instances.

It seems safe to assume that the detection of symmetries in logic programs through reduction to graph automorphism is computationally quite feasible using today's GAP tools such as SAUCY, considering SBASS' runtime.

**Answer Set Enumeration**

Finally, we want to test the impact of symmetry breaking on the number of answer sets. Our study considers instances from the *all-interval series* problem and *graceful graphs*. Recall, the *all-interval series* problem is to find a permutation of the $n$ integers from 0 to $n - 1$ such that the difference of adjacent numbers are also all-different. It has been proposed as a benchmark domain for CP systems by Hoos (1999) and is part of the (CSP library). We modelled the *all-interval series* problem ($AllInt$) as previously described in Example 2.2, using a direct rep-resentation for $n$ integer variables and auxiliary variables to represent the differences between adjacent numbers, and required both sets of variables to be all-different.

Table 6.4: Average time for completed runs in seconds and the number of timeouts on *graceful graph* instances, each shuffled 5 times. The *asterisk denotes instances that have no answer sets.

|  | #gen. | SBASS time | CLASP$_1^\pi$ time | CLASP$_1^\pi$ #t.out | CLASP$_5^\pi$ time | CLASP$_5^\pi$ #t.out | CLASP$^\pi$ time | CLASP$^\pi$ #t.out | CLASP time | CLASP #t.out |
|---|---|---|---|---|---|---|---|---|---|---|
| $DW_3{}^*$ | 5 | 0.02 | 4.24 |  | 1.45 |  | **1.32** |  | 5.40 |  |
| $DW_6$ | 5 | 0.17 | **0.46** |  | 0.56 |  | 1.09 |  | 0.57 |  |
| $DW_8$ | 5 | 0.48 | 28.81 |  | 5.47 |  | 17.11 |  | **4.30** |  |
| $DW_{10}$ | 5 | 1.21 | 191.86 |  | 66.18 |  | **61.59** |  | 27.04 | 2 |
| $DW_{12}$ | 5 | 3.34 | **145.89** |  | 202.18 | 1 | 111.96 | 1 | 112.38 | 4 |
| $K_3P_3$ | 3 | 0.04 | 0.08 |  | 0.08 |  | **0.07** |  | 0.08 |  |
| $K_4P_2$ | 4 | 0.07 | 0.20 |  | **0.10** |  | 0.54 |  | 0.19 |  |
| $K_4P_3$ | 4 | 0.29 | 24.68 |  | 29.06 |  | 198.57 |  | **24.01** |  |
| $K_5P_2$ | 5 | 0.37 | 274.85 | 3 | 334.55 | 3 | **312.56** | 1 | 226.03 | 3 |

Table 6.5: Results on computing all answer sets of selected instances. Runtime and number of solutions are shown.

|  | #gen. | SBASS time | CLASP$_1^\pi$ time | CLASP$_1^\pi$ #sol. | CLASP$_5^\pi$ time | CLASP$_5^\pi$ #sol. | CLASP$^\pi$ time | CLASP$^\pi$ #sol. | CLASP time | CLASP #sol. |
|---|---|---|---|---|---|---|---|---|---|---|
| $AllInt_8$ | 2 | 0.01 | 0.15 | 39 | **0.11** | 15 | 0.17 | 14 | 0.14 | 40 |
| $AllInt_9$ | 2 | 0.01 | 0.78 | 119 | **0.60** | 60 | 0.93 | 40 | 0.77 | 120 |
| $AllInt_{10}$ | 2 | 0.01 | 4.60 | 295 | **3.43** | 148 | 5.69 | 107 | 4.08 | 296 |
| $AllInt_{11}$ | 2 | 0.01 | 23.26 | 647 | **22.82** | 372 | 32.70 | 238 | 24.40 | 648 |
| $AllInt_{12}$ | 2 | 0.01 | 161.90 | 1327 | **147.17** | 862 | 211.27 | 442 | 160.32 | 1328 |
| $DW_4$ | 5 | 0.07 | 282.36 | 9472 | 168.03 | 5152 | **85.65** | 1150 | 314.15 | 11264 |
| $K_3P_3$ | 3 | 0.05 | 229.15 | 5704 | **119.99** | 2836 | 126.25 | 1487 | 268.80 | 6816 |
| $K_4P_2$ | 4 | 0.08 | 119.66 | 1080 | 67.96 | 552 | **27.72** | 146 | 145.13 | 1440 |

As one might expect, we can observe that symmetry breaking significantly compresses the solution-space (see Table 6.5), and therefore, reduces the time necessary for post-processing solutions. Clearly, CLASP$_k^\pi$ discards more solutions (eliminating up to 90 per cent of the solution space) for an increasing number $k$.

Recall that a given problem can be encoded in many equivalent logic programs, and with each different encoding our techniques may detect a different generating set. For instance, we tried symmetry detection and symmetry breaking on logic programs that were preprocessed, i.e., simplified. The key idea of preprocessing logic programs is to identify equivalences among its relevant constituents. These equivalences are then used for building a compact representation of the program (Gebser et al., 2008). Sometimes, we observed significant better results in terms of time and number of answer sets, eliminating up to 95 per cent of the solution space.

## 6.3 Experiments on Constraint Answer Set Programming

Our translational approach to constraint answer set solving has been implemented within the prototypical preprocessor INCA, available at the (Potassco labs suite). It compiles constraint logic programs with first-order variables, function symbols, and aggregates, etc. in linear time and space, such that the logic program can be obtained by a *grounding* process. To evaluate the performance of symmetry breaking in translation-based constraint answer set solving we further modified INCA to handle the *global value precedence* constraint.

Experiments consider INCA in different settings using different decompositions for the *global value precedence* constraint. The model ALL uses our GAC encoding to break all value symmetry. We denote PAIRWISE the method of Law and Lee (2004) which posts *global precedence* constraints between pairwise interchangeable values. The model NONE breaks no symmetry while GENERIC employs SBASS for symmetry breaking in terms of generators. Since INCA is a pure preprocessor, we select the grounder GRINGO (2.0.5) and the ASP solver CLASP (1.3.3) as its backend, using SBASS as middle-ware in case of the GENERIC method. All tests were run on a 2.00 GHz PC under Linux, where each run was limited to 600 s time and 1 GB RAM

**Schur Numbers**

The *Schur number* $S(k)$ is the largest integer $n$ for which the set of integers $1 \ldots n$ can be partitioned into $k$ classes such that the Schur property $x + y = z$ is not satisfied for any triple of integers $(x, y, z)$, where $x, y, z \in 1 \ldots n$ are not necessarily distinct (Guy, 1994). We consider the corresponding decision problem, $S(n, k)$, which asks whether the set of integers $1 \ldots n$ can be partitioned into $k$ classes, all violating the Schur property. This problem has been proposed as a benchmark domain for CP systems by Gent and Walsh (1999) and is part of both the (Asparagus library) and the (CSP library). Furthermore, instances of *Schur's Lemma* formed a benchmark class in the (ASP competition) and the (ASP solver competition). Our basic encoding in the language of the preprocessor INCA defines a constraint variable representing the assignment of each number to a partition, and posts Schur's property:

```
#var $inpart(X) : number(X) = 1..k.

:- $inpart(X) == $inpart(Y), $inpart(X) == $inpart(X+Y).
```

In our configuration, the tool chain in terms of UNIX pipes, `inca | gringo`, generates a logic program with duplicate literals in the body of some rules, whose graph encoding for symmetry detection contains duplicate edges. Since the graph automorphism program SAUCY we employ for symmetry detection rejects graphs with duplicate edges, we developed LNORM (available at (Drescher's research page)), a preprocessor for normalising a logic program, i.e., remove duplicate literals from the body of each rule. In conclusion, the complete tool chain to realise the GENERIC option extends to

```
inca | gringo | lnorm | uniq | sbass | clasp.
```

We follow Law and Lee (2004) and Walsh (2006) and compute all solutions to study the impact of symmetry breaking. The results of our experiments are presented in Table 6.6. Symmetry breaking in terms of generators significantly improves performance, and is itself outperformed by PAIRWISE and ALL. With few interchangeable values, we see similar runtimes using PAIRWISE and ALL. However, ALL is the option of choice when $k$ grows. Our encoding of this global constraint appears therefore to be an efficient and effective mechanism to deal with interchangeable values.

Table 6.6: Results on computing all answer sets of *Schur's Lemma* instances.

|          | NONE    | GENERIC | PAIRWISE | ALL    |
|----------|---------|---------|----------|--------|
| S(13,4)  | 2.71    | 1.13    | 0.13     | 0.11   |
| S(13,5)  | 172.90  | 19.98   | 1.80     | 1.49   |
| S(13,6)  | >600    | 72.74   | 6.93     | 5.65   |
| S(14,4)  | 6.46    | 1.33    | 0.29     | 0.22   |
| S(14,5)  | 527.02  | 68.46   | 6.03     | 4.67   |
| S(14,6)  | >600    | >600    | 30.56    | 27.94  |
| S(15,4)  | 18.42   | 5.81    | 0.70     | 0.59   |
| S(15,5)  | >600    | 257.59  | 22.66    | 18.63  |
| S(15,6)  | >600    | >600    | 153.79   | 125.79 |
| S(16,4)  | 32.29   | 5.76    | 1.35     | 1.10   |
| S(16,5)  | >600    | >600    | 75.23    | 53.87  |
| S(16,6)  | >600    | >600    | >600     | 502.06 |

### Graph Colouring

Recall that a colouring of a graph $(V, E)$ is a a mapping $c$ from $V$ to $\{1, \ldots, k\}$ such that $c(v) \neq c(w)$ for every edge $(v, w) \in E$ with a given number $k$ of colours. Given $k$, the *graph colouring* problem is to determine the existence of a colouring. Random *graph colouring* instances formed a benchmark class in the (ASP competition) and can also be found in the (Asparagus library). We experimented on random graph colouring instances, but restricted ourselves to 3-, 4- and 5-colourings, when we noticed that the relative performance of symmetry breaking increased with each additional colour. For each of the three random graph $k$-colouring experiments we generated 600 instances around the phase transition density with 400, 150, 75 vertices, respectively. To explore the impact of partial PC, we also tried restrictions on the construction of permutation constraints up to the $k$-th atom in a permutation, denoted as GENERIC$_k$.

The data on 3-, 4- and 5- colourings (Table 6.7, and Figures 6.4 and 6.5 which visualise the results better) clearly shows that all symmetry breaking techniques considered in our study are effective, i.e., our proposed methods observably improve the runtime. A detailed analysis of partial PC leaves us with a few clear conclusions. First, but not surprisingly, the first $k$ chain links in the PC constraints prune biggest portions of the search. For the 3-colouring case, the GENERIC$_1$ method reduces the runtime by approximately 50 per cent, and GENERIC$_5$ up to additional 15 per cent. Second, the pruning by longer PC does not pay off (see Table 6.7): The runtime increases with the extra variables introduced by growing $k$, consuming all benefits induced by a smaller search space when using full PC, albeit this effect loses impact with each additional colour. We conclude that GENERIC$_5$ provides a good setting for generic symmetry detection and breaking on our *graph colouring* experiments. It performs slightly better than the PAIRWISE model on the random 3-colouring instances, but worse on the random 4-colouring instances. The data on the 5-colourings clearly shows that our GENERIC method is inferiour to enforcing value precedence through either the method of Law and Lee (2004) or our GAC encoding of the *global value precedence* constraint. For the 3-colouring case, our GAC encoding gives a significantly better improvement compared to the GENERIC and PAIRWISE options. For the 4- and 5-colouring case the same conclusion can be drawn, albeit less convincing. An overall conclusion from our graph colouring experiments must undoubtedly be that breaking value symmetry enforcing some form of *value precedence* is most effective. Given that our encoding of the *global value precedence* constraint is rather simple, we are surprised that no team participating in the second answer set programming competition made use of it (see encodings of (ASP competition) participants).

Table 6.7: Runtime results in seconds on random 3-colouring instances.

| density | NONE | GENERIC$_1$ | GENERIC$_5$ | GENERIC$_{50}$ | GENERIC | PAIRWISE | ALL |
|---|---|---|---|---|---|---|---|
| 2.25 | 21.60 | 12.42 | 10.30 | 11.18 | 20.07 | 13.09 | 0.03 |
| 2.30 | 23.02 | 13.88 | 11.72 | 13.08 | 23.50 | 13.62 | 0.03 |
| 2.35 | 30.06 | 17.09 | 14.51 | 16.22 | 28.40 | 17.11 | 0.58 |
| 2.40 | 29.41 | 14.64 | 11.95 | 13.46 | 22.55 | 14.42 | 0.93 |
| 2.45 | 21.57 | 10.74 | 8.32 | 9.18 | 15.49 | 10.43 | 1.14 |
| 2.50 | 15.45 | 8.20 | 6.50 | 7.19 | 12.84 | 7.96 | 1.32 |

Table 6.8: Runtime results in seconds on random 3-colouring instances.

| density | NONE | ALL$_1$ | ALL$_5$ | ALL$_{20}$ | ALL$_{50}$ | ALL$_{100}$ | ALL$_{200}$ | ALL |
|---|---|---|---|---|---|---|---|---|
| 2.25 | 21.60 | 12.29 | 8.20 | 6.72 | 2.47 | 2.33 | 0.35 | 0.03 |
| 2.30 | 23.02 | 13.55 | 10.73 | 8.93 | 6.70 | 4.32 | 0.95 | 0.03 |
| 2.35 | 30.06 | 16.90 | 13.67 | 12.62 | 8.49 | 4.96 | 2.72 | 0.58 |
| 2.40 | 29.41 | 14.89 | 11.65 | 9.93 | 8.66 | 6.58 | 3.33 | 0.93 |
| 2.45 | 21.57 | 10.48 | 7.73 | 7.23 | 6.02 | 4.46 | 2.54 | 1.14 |
| 2.50 | 15.45 | 8.24 | 6.39 | 6.32 | 4.25 | 3.97 | 3.42 | 1.32 |

To explore the impact of partial symmetry breaking, we also tried restrictions on the construction of the *global value precedence* constraint up to the $k$-th variable in its scope, denoted as ALL$_k$. Our results are shown in Table 6.8 and reveal that – opposed to our observation regarding partial PC – breaking all value symmetries is worth the extra variables introduced in our GAC encoding. Clearly, the more variables in the scope of the *global value precedence* constraint the better the runtime.

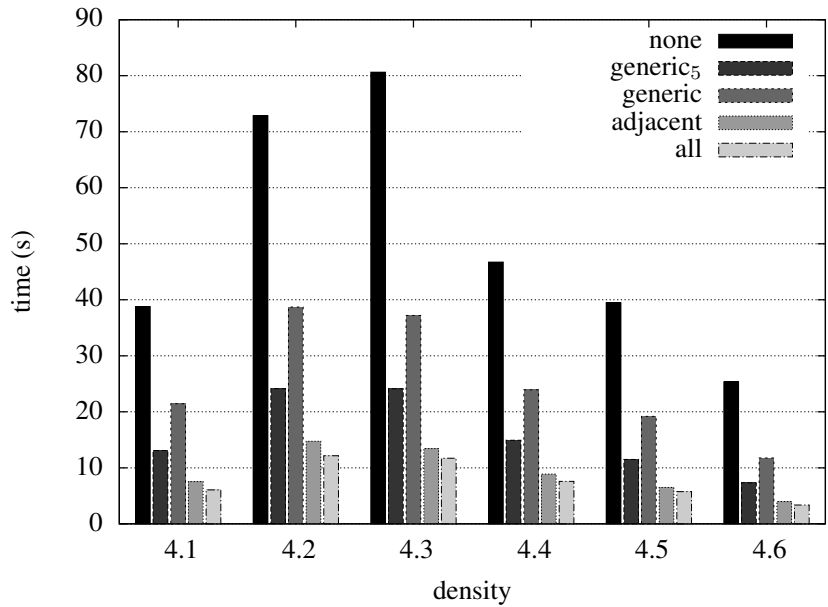

Figure 6.4: Histogram of the average time required by five symmetry breaking approaches to solve random 4-colouring instances near the phase transition density.

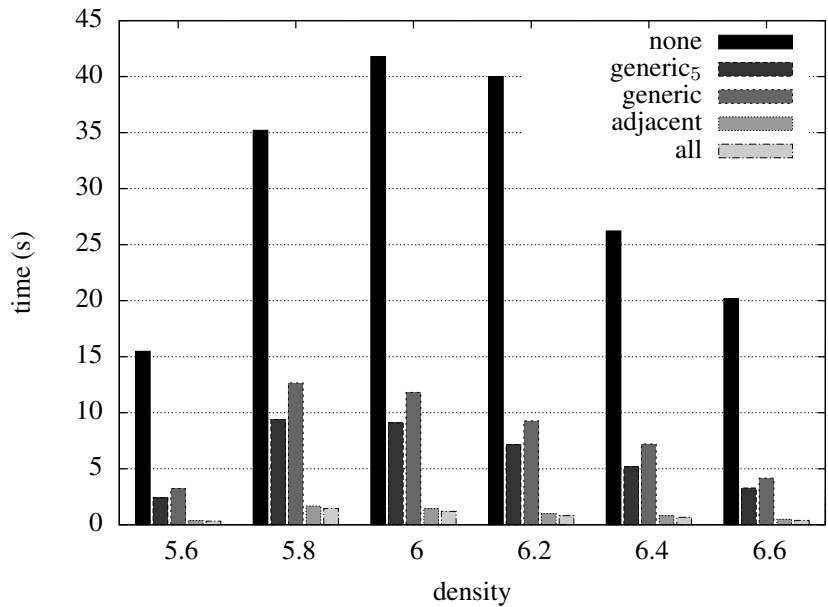

Figure 6.5: Histogram of the average time required by five symmetry breaking approaches to solve random 5-colouring instances near the phase transition density.

Table 6.9: Average time for completed runs in seconds and the number of timeouts on several MCS topologies, 10 random instances each.

|   | $n$ | DMCS time | #t.out | DMCS$^\pi$ time | #t.out | DMCSOPT time | #t.out | DMCSOPT$^\pi$ time | #t.out |
|---|---|---|---|---|---|---|---|---|---|
| D | 10 | 1.90 |   | 0.46 |   | 0.54 |   | **0.35** |   |
|   | 13 | 62.12 | 4 | 32.21 | 2 | 1.38 |   | **0.98** |   |
|   | 25 | — | 10 | — | 10 | 16.12 |   | **11.72** |   |
|   | 31 | — | 10 | — | 10 | 84.02 | 1 | **58.95** |   |
| H | 9 | 7.54 |   | 1.89 |   | 0.33 |   | **0.20** |   |
|   | 13 | 88.85 | 6 | 63.98 | 2 | 0.60 |   | **0.35** |   |
|   | 41 | — | 10 | — | 10 | 1.38 |   | **0.95** |   |
|   | 101 | — | 10 | — | 10 | 5.48 |   | **3.58** |   |
| R | 10 | 0.36 |   | 0.26 |   | 0.15 |   | **0.12** |   |
|   | 13 | 22.41 | 1 | 5.11 |   | 0.19 |   | **0.16** |   |
| Z | 10 | 6.80 |   | 3.24 |   | 0.62 |   | **0.37** |   |
|   | 13 | 57.58 | 3 | 42.93 | 3 | 1.03 |   | **0.68** |   |
|   | 70 | — | 10 | — | 10 | 18.87 |   | **9.98** |   |
|   | 151 | — | 10 | — | 10 | 51.10 |   | **30.15** |   |

## 6.4 Experiments on Distributed Nonmonotonic Multi-Context Systems

Our approach to symmetry breaking for distributed nonmonotonic multi-context systems with ASP logics has been prototypically implemented within a modification of the DMCS system, and its optimized version DMCSOPT (Bairakdar et al., 2010). In contrast to DMCS, DMCSOPT exploits the topology of an MCS, that is the graph where contexts are nodes and import relations define edges, using decomposition techniques and minimises communication between contexts by projecting partial equilibria to relevant atoms. Both systems are available at the (DMCS website). We restrict experiments on MCSs to local symmetry breaking in terms of irredundant generators, and leave unrestricted symmetry breaking, i.e., an exhaustive modification of Bairakdar et al.'s system, to future work. We compare the average response time and the number of solutions under symmetry breaking, denoted as DMCS$^\pi$ and DMCSOPT$^\pi$, on benchmarks versus direct application of the respective systems. All tests were run on a 2 × 1.80 GHz PC under Linux, where each run was limited to 180 s time.

Our benchmarks stem from (Bairakdar et al., 2010) and include random MCSs with various fixed topologies that should resemble the context dependencies of realistic scenarios. In particular, experiments consider MCS instances with ordinary (D) and zig-zag (Z) diamond stack, house stack (H), and ring (R). A diamond stack combines multiple diamonds in a row, where ordinary diamonds (in contrast to zig-zag diamonds) have no connection between the two middle contexts. A house consists of 5 nodes with 6 edges such that the ridge context has directed edges to the two middle contexts, which form with the two base contexts a cycle with 4 edges. House stacks are subsequently built up by using the basement nodes as ridges for the next houses.

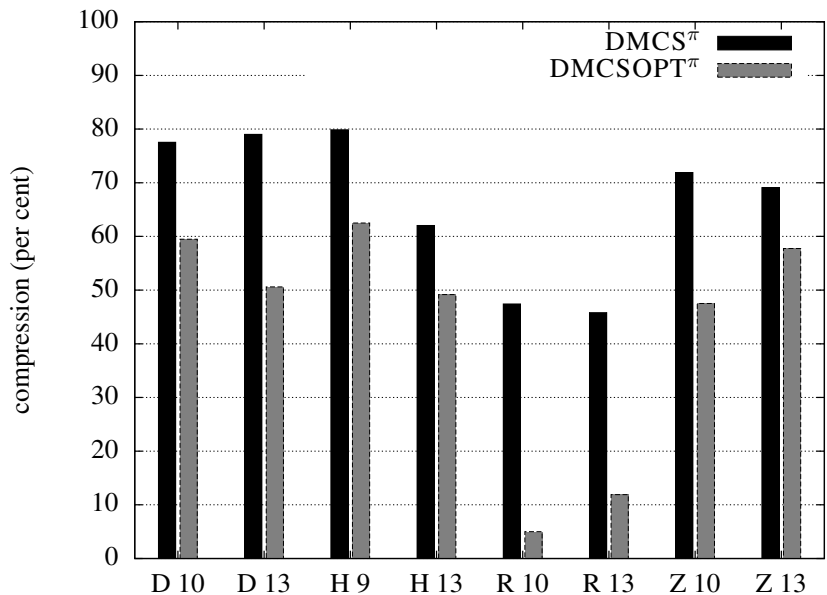

Figure 6.6: Histogram of the average compression of the solution space achieved by symmetry breaking on MCS.

Table 6.9 shows some experimental results on calculating (projected) partial equilibria w.r.t. a randomly selected context of MSC with $n$ contexts, where $n$ varies between 9 and 151. Each context has an alphabet of 10 atoms, exports at most 5 atoms to other contexts, and has a maximum of 5 bridge rules with at most 2 bridge literals. First, we confirm the results of Bairakdar et al., i.e., DMCSOPT can handle larger sizes of MCSs more efficient than DMCS. Second, evaluating the MCS instances with symmetry breaking compared to the direct application of either DMCS or DMCSOPT yields improvements in response time throughout all tested topologies. In fact, symmetry breaking leads to better runtimes on all instances without exception, and in some cases, returns solutions to problems which are otherwise intractable within the given time. Figure 6.6 presents the average compression of the solution space achieved by symmetry breaking. While the results for DMCS$^\pi$ range between 45 and 80 per cent, the impact of symmetry breaking within DMCSOPT on the number of solutions varies between 5 and 65 per cent. We explain the latter with the restriction of DMCSOPT to relevant atoms defined by the calling context.

As a remark, Pólya (1937), Erdős and Rényi (1963) proved that a random graph on $n$ vertices has no symmetries with probability $1 - \binom{n}{2} 2^{-n-2}(1+o(1))$ (c.f. Babai, 1995). This claim can be extended to random MCSs. Since structured instances may have richer symmetries, we expect even more drastic impact of symmetry-breaking on real-world applications.

# 7 Conclusions

Our work addresses solving combinatorial problems in ASP whose difficulty arise from symmetries and redundant search caused by them.

We have shown a reduction of symmetry detection to a graph automorphism problem which allows to extract symmetries of a logic program from the symmetries of the constructed coloured graph. Our techniques are formulated as a completely automated flow that (1) starts with a logic program, (2) detects all of its symmetries within a very general class, including all permutations that do not change the logic program, (3) represents all symmetries implicitly and always with exponential compression in terms of irredundant group generators, and (4) constructs a linear-sized symmetry-breaking constraint that does not affect existence of answer sets. This flow does not require source code modifications in ASP solvers. We successfully validated our implementation with CLASP and SMODELS (SMODELS results are not included in this thesis). Experiments indicate that breaking just the symmetries in a generating set is an efficient and effective way to deal with large numbers of symmetries. In many cases, our techniques achieved significant pruning of the search space and yield solutions to problems which are otherwise intractable. We also observed a significant compression of the solution space which makes symmetry breaking attractive whenever all answer sets have to be post-processed. However, we stress that the proposed flow may not be useful on ASP instances that are easy, or do not have symmetries. Many ASP benchmarks in (Asparagus library) have large numbers of symmetries, but can be solved so quickly that the symmetry detection and breaking overhead is not justified.

We have applied our methods to constraint answer set programming within constraint answer set solving. In particular, we have formulated a translation-based approach to constraint answer set solving which allows for the direct application of our symmetry detection and symmetry-breaking techniques. Although our experiments suggest that symmetry detection seems to be tractable, it is often reasonable to assume that the symmetries for a problem are known. For particular symmetries, there are more efficient breaking methods, for instance, the *global value precedence* constraint which we have decomposed into ASP, such that the unit-propagation of an ASP solver enforces domain consistency on the original constraint. An empirical analysis complements theoretical results and has shown that our decomposition superior to both, our generic method and Law and Lee's approach.

We have also extended our methods to distributed answer set programming in the framework of multi-context systems. In particular, we have presented a basic distributed algorithm for computing all (partial) symmetries of an MCS, and also carried over symmetry-breaking constraints. The utility of our approach is not clear at this moment, e.g. symmetries are not represented efficiently, for instance, in terms of generators. However, we have conducted experiments on distributed symmetry breaking and got promising results.

**Future Work**

We want to put forward constraint answer set programming as a novel approach to constraint (logic) programming. Therefore, we (1) investigate efficient encodings of propagation algorithms in answer set programming, (2) study the integration of techniques from constraint processing into answer set programming engines, and (3) define a modelling language for constraint logic programming under answer set semantics, that can be accepted by the scientific community. Furthermore, we (4) want to implement our techniques in state-of-the-art systems. In particular, future work concerns, but is not limited to, the integration of further constraints useful in constraint answer set programming. We are interested in decompositions of constraints using the full range of propagators available in state-of-the-art ASP systems, and where necessary, extending ASP solving by further useful algorithms that make constraint answer set programming more powerful. Motivated by award-winning application of the ASP solver CLASP as a pseudo-Boolean solver (PB competition, 2009) and a SAT solver (SAT competition), we conjecture that an ASP system can perform similar in traditional constraint solver competitions.

We also belief that multi-context systems provide foundations to distributed answer set programming. In this regard, we want to investigate techniques to efficiently represent partial symmetries, e.g., in terms of irredundant generators, and plan to implement our approach based on Bairakdar et al.'s system.